\documentclass{emulateapj}
\usepackage{natbib}

\usepackage{epsfig}
\usepackage{subfigure}
\usepackage{graphics}

\newcommand{\chan}{{\it Chandra}}

\newcommand{\kms}{$\,\rm{km\,s^{-1}}$}

\shorttitle{Anatomy of Abell~1201}
\shortauthors{Owers et al.}
\begin{document}
\title{Abell~1201: The anatomy of a cold front cluster from combined optical and X-ray data}
\author{Matt S. Owers\altaffilmark{1}, Paul E.J. Nulsen\altaffilmark{2}, Warrick J. Couch\altaffilmark{3}, Maxim Markevitch\altaffilmark{2}}
\author {Gregory B. Poole\altaffilmark{3}}
\altaffiltext{1}{School of Physics, University of New South Wales, Sydney, NSW 2052, Australia; mowers@phys.unsw.edu.au}
\altaffiltext{2}{Harvard Smithsonian Center for Astrophysics, 60 Garden Street, Cambridge, MA 02138, USA}
\altaffiltext{3}{Center for Astrophysics and Supercomputing, Swinburne University of Technology, Hawthorn, VIC 3122, Australia}

\begin{abstract}

We present a combined X-ray and optical analysis of the cold front cluster Abell~1201 using archival \chan\, data and
multi-object spectroscopy taken with the 3.9m Anglo Australian and 6.5m Multiple Mirror Telescopes. This paper represents 
the first in a series presenting a study of a sample of cold front clusters selected from the \chan\, archives with the 
aim of relating cold fronts to merger activity, understanding the dynamics of mergers and their effect on the cluster 
constituents. The \chan\, X-ray imagery of Abell~1201 reveals two conspicuous surface brightness discontinuities, which 
are shown to be cold fronts, and a remnant core structure. Temperature maps reveal a complex multi-phase temperature 
structure with regions of hot gas interspersed with fingers of cold gas. Our optical analysis is based on
a sample of 321 confirmed members, whose mean redshift is $z=0.1673\pm 0.0002$ and velocity dispersion is $778\pm36$\kms. 
We search for dynamical substructure and find clear evidence for multiple localized velocity substructures coincident 
with over-densities in the galaxy surface density. Most notably, we find structure coincident with the remnant X-ray core.
Despite the clear evidence for dynamical activity, we find the peculiar velocity distribution does not deviate 
significantly from Gaussian. We apply two-body dynamical analyses in order to assess which of the substructures are bound,
and thus dynamically important in terms of the cluster merger history. We propose that the cold fronts in Abell~1201 are 
a consequence of its merger with a smaller subunit, which has induced gas motions that gave rise to `sloshing' cold 
fronts. Abell~1201 illustrates the value of combining multi-wavelength data and multiple substructure detection 
techniques when attempting to ascertain the dynamical state of a cluster. 
\end{abstract}


\section{Introduction}

Within the current cosmological paradigm, large scale structure in the Universe is expected to form in a hierarchical 
manner. This ``bottom up'' formation scenario culminates with the formation of clusters of galaxies, which are the largest
and most massive virialized objects in the Universe. At the current epoch, a high fraction of clusters are still growing 
through the infall of matter, much of which is funneled through the surrounding spider-web like filamentary structures. 
The most extreme growth event occurs when two clusters of roughly equal mass merge, the merging process being one of the 
most energetic in the Universe, releasing around $10^{64}$ ergs of gravitational binding energy. Some $10\%$ of this is 
dissipated via mechanisms such as shock and adiabatic heating of the intracluster medium (ICM), acceleration of 
relativistic particles, and generation of peculiar velocities and turbulence in the ICM \citep[for a review of these 
physical processes see][]{sarazin2002}.

In this context, the new generation of X-ray observatories (XMM-Newton and \chan) have provided a wealth of new 
information on the ICM and hence new insights into this cluster merger phenomenon. One of the first interesting 
discoveries -- attributed to the excellent spatial resolution and sensitivity of \chan\ --  was the observation of 
extended `edge' features in the X-ray surface brightness maps obtained for Abell~2142 \citep{markevitch2000} and 
Abell~3667 \citep{vikhlinin2001}. The sharp discontinuity in surface brightness seen in Abell~3667 had been observed 
previously with {\it ROSAT}, and was interpreted as a shock front \citep{markevitch1999}. However, the \chan\, 
observations showed that in both Abell~2142 and Abell~3667,  the temperature of the gas on the brighter (denser) side of 
the discontinuity was {\it colder} than the temperature of the less dense downstream gas -- opposite to expectations for 
a shock. Furthermore, the temperature and density profiles across the discontinuities showed, in combination, that the 
pressure is roughly continuous, again inconsistent with a shock front. Thus, the observed edges appear to be contact 
discontinuities between cool, dense, low entropy gas and hotter, diffuse ambient ICM, leading them to be dubbed 
``cold fronts''. Initial explanations attributed the low entropy gas to the remnant cooling core of a merging sub-cluster
\citep{markevitch2000}.

As the number of clusters observed to have cold front features has increased, it has become clear that the remnant core 
scenario is not the correct interpretation in all cases, since a number of cold fronts are observed to exist in clusters 
with an otherwise relaxed X-ray morphology \citep{mazzotta2001b,mazzotta2001,markevitch2001,markevitch2007}. One such 
example is Abell~1795, the observations of which led \citet{markevitch2001} to propose an alternative scenario whereby 
the `sloshing' of the cooling core within the stationary gravitational potential well causes cool central gas to be 
displaced and a cold front formed where it comes into contact with higher entropy gas at larger radii. 

Hydrodynamic simulations have an integral role in aiding the interpretation of these cluster 
X-ray observations and further understanding the underlying physics. Both \citet{ascasibar2006} and \citet{poole2006} 
found a plethora of transient cold-front like phenomena associated with sub-cluster gas arising during their simulated 
mergers, including the classic ram-pressure stripped remnant sub-cluster core preceded by a merger shock. 
\citet{tittley2005} explored the possibility of forming a cold front through gas sloshing, finding it is possible to 
produce similar edges to those observed through core oscillations, although in their model they find the gravitational 
potential well oscillates due to the motion of the dark matter, inducing motion in the gas. \citet{churazov2003} and 
\citet{fujita2004} also showed that gas oscillations can be induced when the cluster core is displaced from the potential 
well by a weak shock or acoustic wave which has passed through the cluster center. \citet{ascasibar2006} found that 
oscillation of the central gas and dark matter can easily be induced by infalling sub-clusters, and can even be induced
due to the infall of a dark matter only sub-cluster. The gas 
core decouples from the dark matter when a rapid change in the direction of oscillation due to the core passage of the 
dark matter sub-cluster causes a change in the ram pressure felt by the gas, displacing it from the dark matter. The gas 
sloshing then occurs when the displaced gas falls back toward the potential minimum, generating edges at the turn-around 
point of each oscillation. 

Whatever the mechanism, it appears the existence of a cold front can generally be interpreted as strong evidence of a 
system which is in the process of, or has recently undergone a merger. What has been lacking to date is a systematic 
study of the relationship between cold fronts and other dynamical indicators of cluster merger activity using 
multi-wavelength observations, which can be incorporated into detailed models in order to garner a complete understanding 
of how a major merger impacts the different cluster mass components. To address this issue, we have conducted a search
of the \chan\, archives and selected a sample of clusters exhibiting robust examples of cold fronts (the 
selection criteria and sample will be presented in a forthcoming paper; Owers et al. 2008, in prep.) for optical 
multiple-object spectra (MOS) and radio follow up observations. This paper represents the first of a series where we 
aim to show a relationship between cold fronts and merger activity through detection of substructure at optical 
wavelengths. Subsequent papers will relate the star formation/radio properties of the galaxies and 
large scale diffuse radio halo/relic emission associated with cluster mergers \citep{giovannini2002}.  

Detection of substructure within clusters using optical data has a long history. Initially, galaxy surface density 
contours were used to search for projected galaxy concentrations within clusters \citep{geller1982}. Advances in 
multi-object spectroscopy have allowed simultaneous observations of tens, and now hundreds of galaxy spectra within a 
field, meaning large samples of spectroscopically confirmed cluster members can now be compiled. Radial velocity 
information aids in eliminating the projection effects inherent in galaxy surface density contour methods, and a number 
of methods have been proposed which use the combination of galaxy radial velocity and spatial information to develop a 
statistic for the robust detection of substructures within a cluster \citep[eg. see][]{dressler1988, west1990, colless1996,
girardi2002}. Since different statistics are sensitive to different types of substructure \citep{pinkney1996}, 
the use of a combination of a number of different statistical methods is essential in diagnosing the dynamical activity
in a cluster. Combining these substructure detection methods with data at X-ray wavelengths has proven to be a powerful
tool in diagnosing cluster merger scenarios and is essential for disentangling the complex histories of merging systems
\citep[eg.][]{boschin2004,girardi2006,barrena2007,carrasco2007,maurogordato2008}.

Here, in this first paper, we present a detailed optical and X-ray analysis of the cluster Abell~1201. The (previously 
unpublished) \chan\, observations of this cluster reveal two sharp surface brightness discontinuities, both of which are 
cold fronts, and also an offset core of X-ray emission. Based on comparisons of the simulations of \citet{poole2006} and
\citet{ascasibar2006} and these observations, Abell~1201 appears to be an excellent example of cold fronts generated by 
core gas motions caused by a gravitational perturbation in the form of a merging subcluster. Thus, Abell~1201 provides a 
unique opportunity to test this scenario using combined optical and X-ray analyses. 

Abell~1201 is a richness class 2 cluster \citep{abell1989} at 
moderate redshift z=0.168 \citep{struble1999} with an X-ray luminosity 
$\rm{L}_{\rm x}=2.4~\times~10^{44} \rm{ergs}\, \rm{s}^{-1}$ \citep{bohringer2000}. The analysis presented is based on 
comprehensive optical spectroscopy obtained using both AAOmega on the Anglo-Australian Telescope (AAT) and 
Hectospec on the Multiple-Mirror Telescope (MMT) combined with \chan\, archival X-ray data. The structure of the paper 
is as follows: In \S\ref{xrayanal} we present the reduction and analysis of the \chan\, data. In \S\ref{opticalanal} we 
present the new MOS optical observations and analysis. In \S\ref{mergescen} we present a scenario for the merger history 
and formation of the cold fronts in Abell~1201. In \S\ref{discussion} we summarize our results.

Throughout the paper, we assume a standard $\Lambda \rm{CDM}$ cosmology where $H_0=70$\kms, $\Omega_m=0.3$ 
and $\Omega_{\Lambda}=0.7$. For this cosmology and at the redshift of the cluster $1\arcsec=2.88$\,kpc.

\section{X-ray Observations and Analysis}\label{xrayanal}

Our X-ray analysis of Abell~1201 uses \chan\, archival data that were taken on 2004 November 3 (ObsId 4216). The 
observations were made with the ACIS-S array with the cluster centered on the back illuminated S3 chip, for a total 
exposure time of 40\,ksec.  

\subsection{Data preparation}\label{data.prep}

The data were reprocessed using the CIAO software package (version 3.4) starting from the standard \chan\, pipeline 
processed level-1 event list. Observation specific bad pixel files containing hot pixel and cosmic ray afterglow 
information were produced and applied, the latest gain files and calibrations were applied and VFAINT mode cleaning was 
used for improved rejection of cosmic ray events. The data were then filtered to include only events with {\it ASCA} 
grades 0, 2, 3, 4 and 6.

Since the backside illuminated chips are prone to flare contamination, we filtered the data for periods of anomalously 
high background. We followed standard procedure and extracted lightcurves from source-free regions to search for flares. 
Cluster emission covers much of the S3 chip, so we extracted a lightcurve in the 2.5-6\,keV range using the S1 chip which 
has very similar flare properties to the S3 chip. The observation suffered from a strong flare, and approximately half the 
exposure time was affected, leaving a useful exposure time of 21.5\,ks. During the spectral analyses presented 
below, this cleaned data is used. For the imaging analyses, including the fitting of surface brightness profiles, the full 
40ks uncleaned exposure is used.

Background data were taken from the blank sky observations appropriate for the epoch of 
observation\footnote{See http://cxc.harvard.edu/contrib/maxim/acisbg/}. The backgrounds are taken from observations with
low Galactic foregrounds and soft X-ray brightness, so we check the soft X-ray flux in the vicinity of Abell~1201
using the {\it ROSAT} all sky R45\footnote{See http://heasarc.gsfc.nasa.gov/docs/tools.html} count rates and confirm they are
consistent with the blank sky background rates. The background and source data were processed 
using the same calibration files, bad pixel files and background filtering, with the backgrounds being reprojected to 
match the observations. 

We check for residual backgrounds in the cleaned data by extracting spectra from the S1 chip and after subtracting 
backgrounds from the blank sky observations, find a small residual background. This residual was modeled by a soft thermal 
MEKAL component with solar abundance and kT=0.27\,keV \citep[similar to that found in][]{markevitch2003} plus a cutoff 
power law, which is not folded through the instrument response, with photon index of -0.15 and exponential cutoff at 
5.6\,keV \citep{markevitch2003}. Including these components, scaled for region size, in the spectral fits performed in
\S~\ref{spectral} makes no significant difference to the results.

\subsection{X-ray images and Surface Brightness Modeling}

\begin{figure*}
\begin{tabular}{cc}
{\includegraphics[angle=0,width=0.48\textwidth]{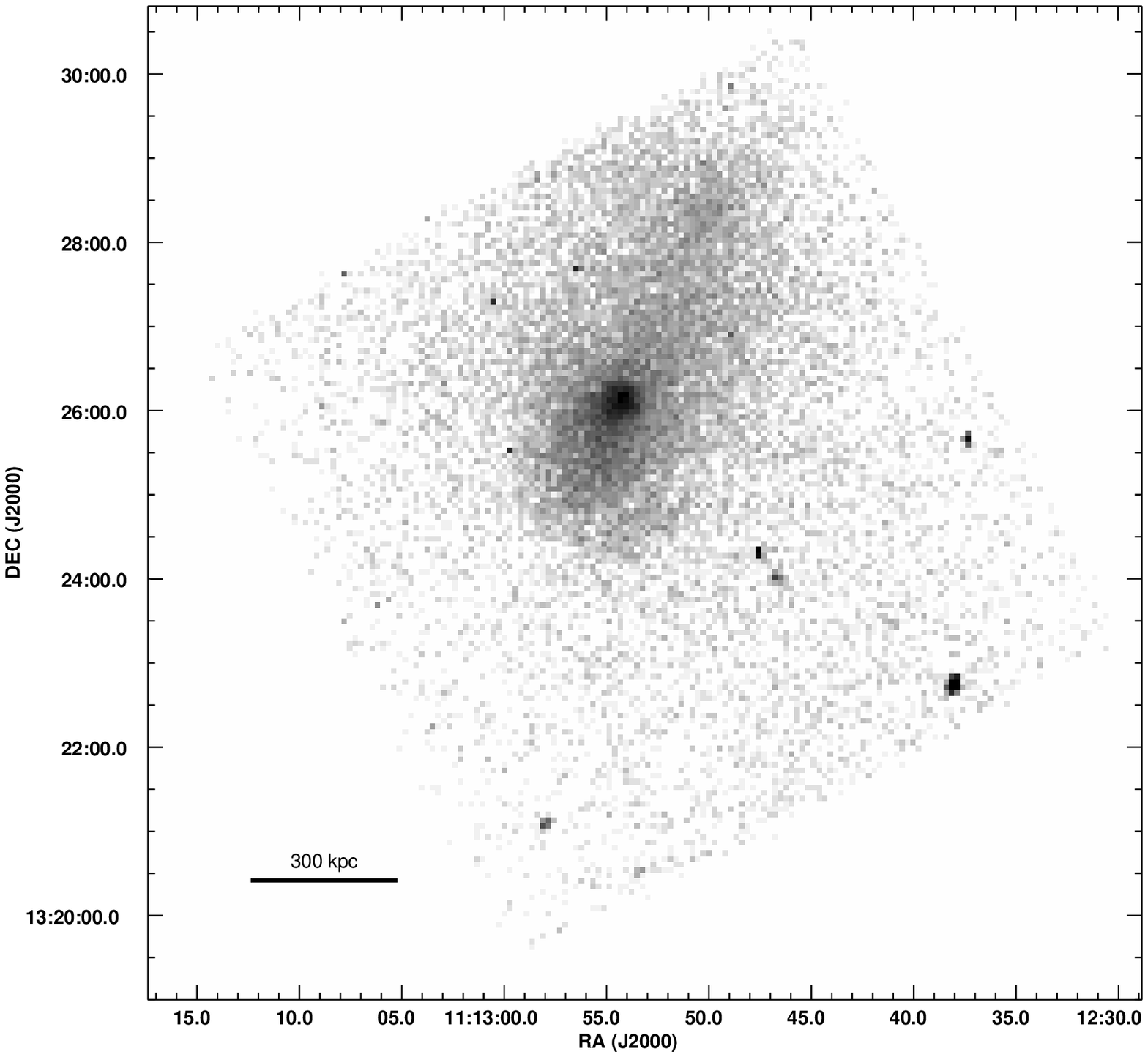}}
{\includegraphics[angle=0,width=0.45\textwidth]{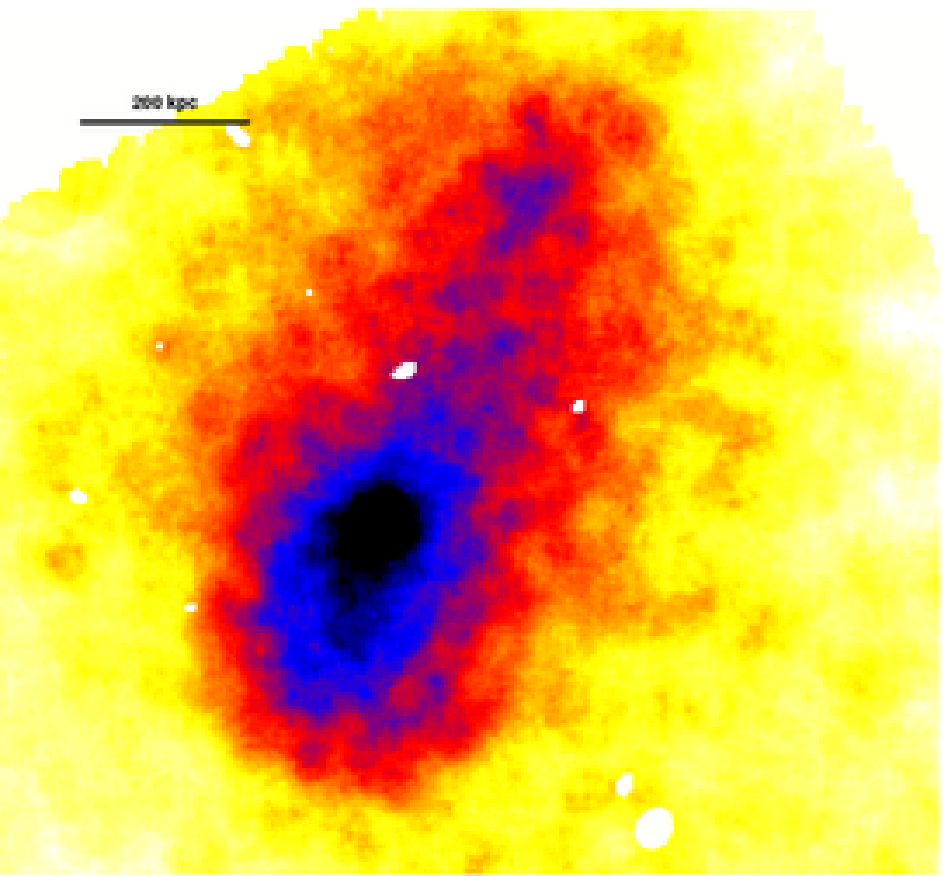}}
\end{tabular}
\caption{{\it Left:} Raw \chan\, ACIS-S3 X-ray image of Abell~1201 in the 0.5-7.0\,keV energy range, binned to 4\arcsec\, 
pixels, including only the S3 chip, using the full 40ks exposure and displayed on a logarithmic scale to enhance the 
diffuse, low surface brightness emission. North is up and east is to the left. {\it Right:} Adaptively
smoothed, exposure corrected close up version of the {\it left} image with point sources removed. The emission appears 
elongated on an axis from the south-east to north-west, and there is a bright central X-ray core. To the north-west there 
is a clump of excess emission and there are two surface brightness discontinuities, one to the south-east and one near 
the core. }
\label{rawchan}
\end{figure*}

\begin{figure}
{\includegraphics[angle=-0,width=.48\textwidth]{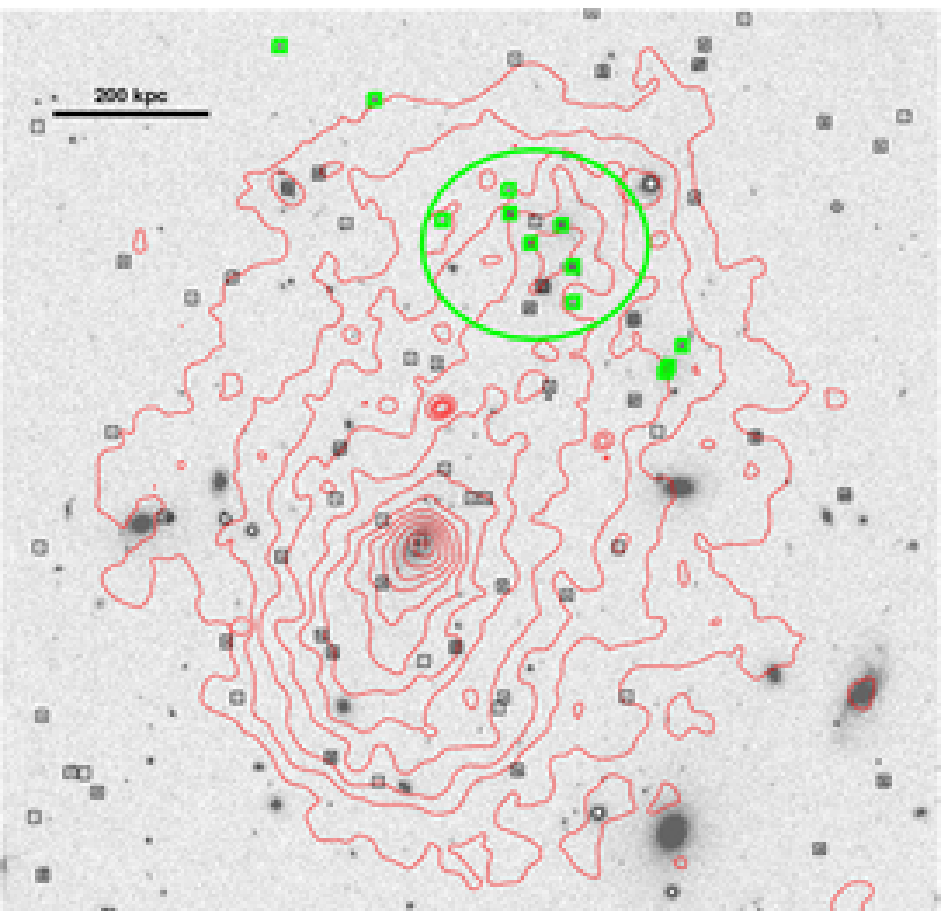}}
\caption{SDSS r-band image of the central region of Abell~1201. The {\it red contours} are the \chan\, X-ray contours 
logarithmically spaced by a factor of 1.3 in the interval 
$1.5 \times 10^{-8}-4.6 \times 10^{-7}$~photons\,cm$^{-2}$\,s$^{-1}$\,pixel$^{-1}$, the {\it black squares} identify 
galaxies which are cluster members, the {\it green squares} show galaxies allocated to KMM1, and the {\it green ellipse} 
shows the 1$\sigma$ region for KMM1 from the KMM analysis (see \S~\ref{3dstructures}). The bright cluster galaxy 
{\it white square} appears to be associated with the bright X-ray core, although is offset by $\sim2.8$\arcsec.}
\label{optxraycont}
\end{figure}

The raw 0.5--7.0\,keV \chan\, image, uncorrected for exposure and showing the entire ACIS-S3 chip, is presented in 
the left panel of Figure~\ref{rawchan}. The right panel of Figure~\ref{rawchan} shows the corresponding exposure 
corrected, adaptively smoothed \chan\, image. The cluster has a bright core, and appears elongated along an axis pointing
to the north-west. Along this axis, there are three distinct features apart from the cool core: There are two surface 
brightness discontinuities, one to the south-east $\sim 300$\,kpc from the core running from position angle (PA; measured
from due west) $240^{\circ}$ to $267^{\circ}$ and one $\sim 50$\,kpc north-west of the core (PA $8^{\circ}$ to 
$120^{\circ}$) and $\sim 430$\,kpc to the north-west there is a faint, diffuse excess clump of emission. We discuss the 
discontinuities further in \S\ref{sb.disc}. In Figure~\ref{optxraycont} we show the SDSS r-band optical image of the 
central regions of Abell~1201 with \chan\, X-ray contours overlaid. The X-ray peak is offset by $\sim2.8$\arcsec\, 
from the brightest cluster galaxy (BCG) and the excess clump of emission approximately coincides with a second 
concentration of galaxies, which will be discussed further in \S\ref{opticalanal} and \S\ref{mergescen}.

\subsubsection{Beta model}\label{betaresid}

In order to emphasize the structures seen in Figure~\ref{rawchan}, we use the method of \citet{neumann1997}. The
surface brightness distribution is fitted with an azimuthally symmetric model and the residuals are inspected for 
significant deviations. We use the {\it Sherpa} fitting package to fit a background 
subtracted double-Beta model of the form
\begin{equation}
S(r)= \sum_{i=1}^2{S_{0,i}\left[1+\left({{r} \over {r_{0,i}}}\right)^2\right]^{-3 \beta +1/2}} +B,
\end{equation}
where $r$ is the radius, measured from an adjustable origin. One Beta model accounts for the excess emission within the 
central $~100$\,kpc and the second models the remaining cluster emission. For fitting, the model is multiplied by the 
exposure map to account for instrumental effects. The best fitting model parameters are presented in 
Table~\ref{betaparams}. 

\begin{figure}
{\includegraphics[angle=-0,width=0.48\textwidth]{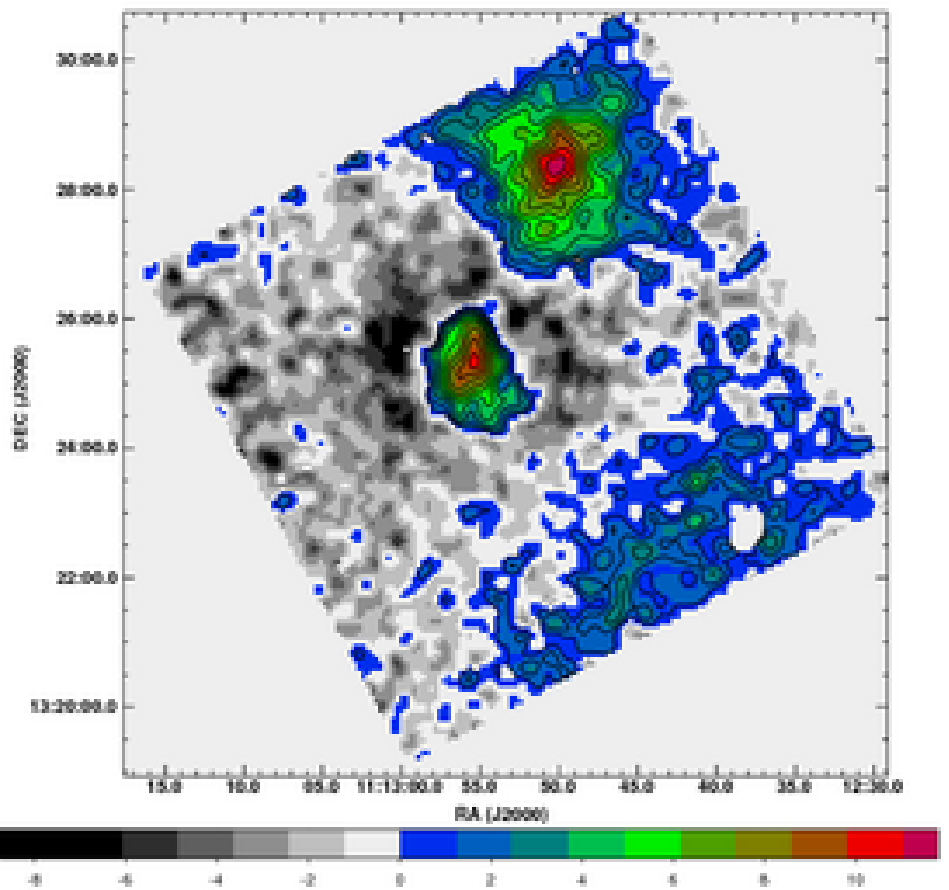}}
\caption{Residual significance map derived from subtracting a smooth double Beta model from the X-ray image and 
calculating the significance (see text). Grey scales show the negative residuals, whilst color scales show the positive 
residuals. The black contours range from $1-12\sigma$  with linear increments of $1\sigma$.}
\label{residsig}
\end{figure}

We obtain residual maps by subtracting the smooth model described above from the entire cluster image. This residual map 
is smoothed with a Gaussian kernel with $\sigma=5.9$\arcsec\, and residual significance is determined using the method of 
\citet{neumann1997}, where Gaussian smoothing and Poissonian statistics are used to determine an error map for the X-ray 
image, and the significance map is the ratio of the residual to error map. Figure~\ref{residsig} shows the residual 
significance map, with the north-west clump detected at greater than $10\sigma$ significance while an excess plume is 
detected in the direction of the south-east discontinuity. There appears to be a significant excess detected in the 
south-west at the bottom of the chip. This can be attributed to excess emission due to the background flare, since the 
excess is much less significant when the same analysis is performed on the cleaned 21.5\,ks exposure. Also 
apparent are the significant negative residuals surrounding the south-east plume and in the outer parts of the chip to 
the west, artifacts of the azimuthally symmetric model.

\begin{deluxetable*}{ccccccc}
\tablecolumns{7}
\tablewidth{0pc}
\tabletypesize{\scriptsize}
\tablecaption{Beta model parameters derived from fitting the surface brightness distribution with a double beta model 
(see text).\label{betaparams}}
\tablehead{\colhead{$x_0, y_0$}&\colhead{$S_{0,1}$} &\colhead{$S_{0,2}$} &\colhead{$\beta$}&\colhead{$r_{0,1}$}&\colhead{$r_{0,2}$}& \colhead{$B$}\\
\colhead{(deg, J2000)} &\colhead{($10^{-7}$)}&\colhead{($10^{-7}$)}&&\colhead{(kpc)}&\colhead{(kpc)}&\colhead{($10^{-9}$)}
}
\startdata
 $11^{\rm h}12^{\rm m}54.3^{\rm s},\,+13^{\circ}26\arcmin06.7\arcsec$& $3.53^{+0.38}_{-0.32}$&$0.44^{+0.03}_{-0.04}$ 
&$0.50^{+0.04}_{-0.04}$&$22^{+4}_{-3}$&
$213^{+22}_{-24}$&$6.56^{+0.67}_{-0.54}$ \\
\enddata
\tablecomments{Units of $S_{0,1},\, S_{0,2}\, {\rm and}\, B$ are $photons/cm^2/s/arcsec^2$. Errors on $x_0$ and 
$y_0$ are $\sim 0.4$\arcsec.}
\end{deluxetable*}

\subsection{Global spectral properties and the nature of the surface brightness discontinuities}\label{spectral}

In this section we derive global properties for Abell~1201, and also analyze the properties of the sharp surface brightness
discontinuities seen in Figure~\ref{rawchan} with the aim of determining the nature of the fronts. Table~\ref{gasprops}
summarizes the various best fitting spectral parameters measured in the following sections.

\begin{deluxetable}{ccc}
\tabletypesize{\scriptsize}
\tablecolumns{3}
\tablewidth{0pc}
\tablecaption{Summary of best fitting spectral parameters for several regions.\label{gasprops}}
\tablehead {\colhead{Region} & \colhead{kT} & \colhead{Abundance}\\
& \colhead{(keV)} & \colhead{($Z$)}}
\startdata
Global & $5.3\pm0.3$&$0.34\pm0.10$ \\
Inside south-east front & $3.6^{+1.0}_{-0.7}$& \nodata\\
Outside south-east front & $5.7^{+3.7}_{-1.7}$& \nodata\\
Inside north-west front & $3.2^{+1.0}_{-0.7}$& \nodata\\
Outside north-west front & $5.2^{+1.5}_{-1.0}$& \nodata\\
\enddata
\end{deluxetable}

\subsubsection{Global Temperature}\label{global.temp}

\begin{figure}
{\includegraphics[angle=0,width=0.48\textwidth]{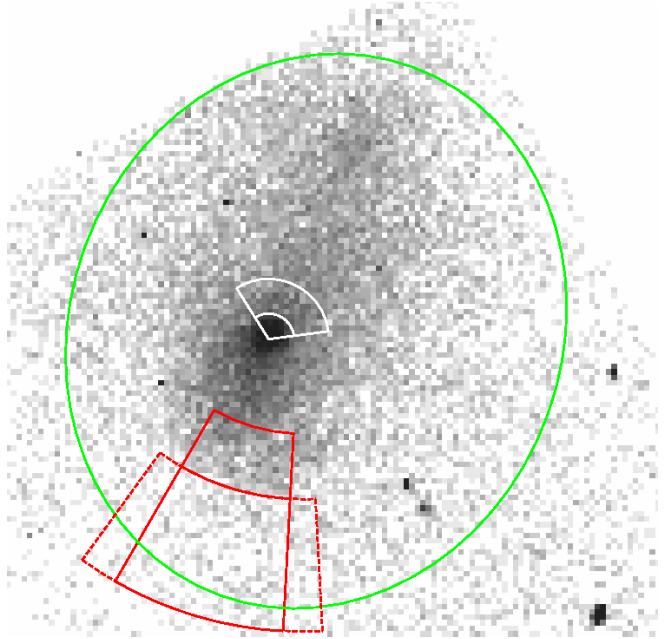}}
\caption{Close up of the raw \chan\, image with regions showing where surface brightness profiles were fitted, and where
spectra were extracted for temperature measurements. The large green ellipse shows the region where spectra were extracted
for global temperature and abundance measurements. The surface brightness profiles shown in Figure~\ref{sbprof} are taken
from the solid red and white sectors. The boundaries indicate the fronts, and delineate regions where the temperature
inside and outside the fronts are measured. The dashed red sector indicates the slightly larger opening angle used in 
measuring the temperature outside the south-east front.}
\label{bin8tempregs}
\end{figure}

We derived a global temperature and metallicity for Abell~1201 within an elliptical region covering the majority of the 
cluster emission (see Figure~\ref{bin8tempregs}). The spectrum was extracted using the CIAO {\sf dmextract} tool, and 
point sources detected with {\sf wavdetect} excised. The CIAO tool {\sf mkwarf} was used to create the auxiliary response 
file (ARF), which accounts for spatial variations in quantum efficiency (QE), effective area and the buildup of contaminant
on the filter windows. The observation was performed at $-120^\circ$C, so the CIAO tool {\sf mkacisrmf} was used to create 
redistribution matrix files (RMF). The ARF and RMF calculated within a region are weighted by the number of events in the 
0.5-2\,keV range within the region. 

The unbinned spectra were fitted in the 0.5-9.8\,keV range using the MEKAL \citep{kaastra1992,1995ApJ...438L.115L} 
and WABS models within the XSPEC package \citep{1996ASPC..101...17A}, minimizing the Cash statistic. The MEKAL
component models a hot, diffuse, single temperature plasma, and is multiplied by WABS, a photoelectric absorption 
model accounting for Galactic absorption, where the neutral hydrogen column density is fixed to the Galactic value of 
$N_H=1.61 \times 10^{20}$ \citep{1990ARA&A..28..215D}. The best fitting values for the temperatures and abundance are 
$kT=5.3^{+0.3}_{-0.3}$ and $Z=0.34^{+0.10}_{-0.10}$ where the errors are $90\%$ confidence limits. We used the LUMIN
function to determine the integrated unabsorbed X-ray luminosity from the best fitting model in the energy range 
0.5-7.0\,keV, and obtained $L_{\rm X}(0.5-7.0\,{\rm keV})=3.7 \times 10^{44}$ $\rm erg\,s^{-1}$ within the 
$r \simeq 550$\,kpc elliptical aperture shown in Figure~\ref{bin8tempregs}.

\subsubsection{Surface Brightness Discontinuities}\label{sb.disc}

\begin{figure*}
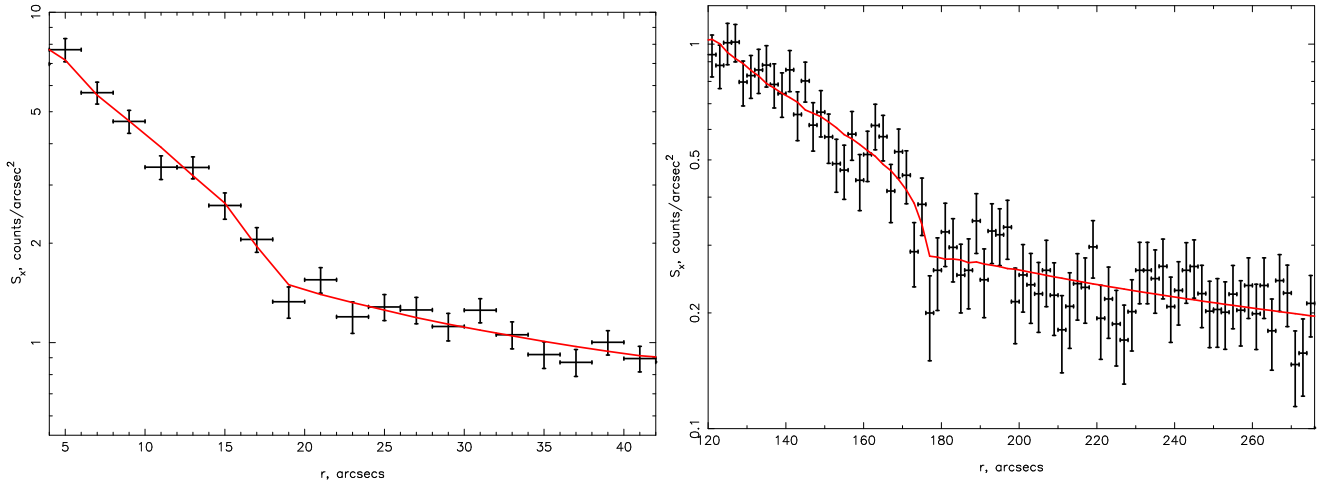

\begin{tabular}{cc}
{\includegraphics[angle=-90,width=0.48\textwidth]{inner.cf.ps}}
{\includegraphics[angle=-90,width=0.48\textwidth]{outer.cf.ps}}
\end{tabular}
\caption{{\it Left:} Surface brightness profile across the north-west discontinuity measured in a sector from PA 
$7^{\circ}-122^{\circ}$ centered at the center of curvature of the front (RA=168.2268, DEC=13.4344).{\it Right:} Surface 
brightness profile across the south-east discontinuity  measured in a sector from PA $240^{\circ}-267^{\circ}$ centered at 
the center of curvature of the front (RA=168.2197, DEC=13.4509). The red curves show the surface brightness profile 
for the best fitting density model.}
\label{sbprof}
\end{figure*}

To characterize the surface brightness discontinuities as either shock or cold fronts, we fitted the surface 
brightness profiles across them with a density model, and also measured the temperature difference across the 
discontinuities. The density model consists of a broken power law function such that 
\begin{equation}
n_e(r)=\left\{
\begin{array}{ll}
  n_{e,1}\left({{r}\over{R_f}}\right)^{-\alpha_1}, &  r<R_f,\\
  n_{e,2}\left({{r}\over{R_f}}\right)^{-\alpha_2}, &  r>R_f,
\end{array}
\right.
\label{density}
\end{equation}
where $R_f$ is the radius at which the discontinuity occurs, $n_{e,1}$ and $n_{e,2}$ are the densities of the inner and 
outer gas at $R_f$, respectively, and the spheroidal radius, $r$, is defined by $r^2=\varpi^2+{\epsilon_\zeta}^2 \zeta^2$. 
Here $\zeta$ is the coordinate along our line of sight and the elliptical radius, $\varpi$, is defined by
\begin{equation}
\varpi^2={{\epsilon^2[x {\rm cos}\theta+y {\rm sin}\theta]^2+[y {\rm cos}\theta-x {\rm sin}\theta)]^2} \over {\epsilon^2}},
\end{equation}
where $x$ and $y$ are cartesian coordinates centered at the center of curvature of the discontinuity and $\theta$ 
determines the orientation of the front on the plane of the sky. We assumed the gas density distributions inside and 
outside the front have the same ellipticities and have rotational symmetry so that $\epsilon_\zeta=1/\epsilon$. The model 
is incorporated into {\it Sherpa} where we fitted the surface brightness across the discontinuity in the sectors shown in 
Figure~\ref{bin8tempregs} using the full 40\,ks exposure, in the energy range 0.5-7\,keV, and also add to the model a 
constant background component which was measured in \S~\ref{betaresid}. We present the surface brightness distribution, 
along with the best fitting density models, in Figure~\ref{sbprof}. From the amplitudes of the best fitting surface 
brightness model, $A_1$ and $A_2$, we derived a density jump of $\sqrt{A_1/A_2}={n_{e,1}/ n_{e,2}}=2.13^{+0.56}_{-0.39}$ 
for the south-east discontinuity and ${n_{e,1}/ n_{e,2}}=1.75^{+0.47}_{-0.34}$ for the north-west discontinuity. The 
confidence range for the density jump was computed from the extremes of the $90\%$ confidence ranges for $A_1$ and $A_2$, 
assuming a single parameter of interest.  We note that if the density jump is the only parameter of interest, this  
overestimates its confidence range.

The temperature profile across the front is an essential diagnostic tool. Ideally, to minimize projection effects  
we would like to measure the deprojected temperature profile across the fronts, however the limited number of photons 
available do not allow this. Instead, we extracted spectra from two regions for each front, one on the bright side of the
front (inside) and one on the faint side (outside). The regions from which we extracted spectra for both the south-east and north-west 
fronts are shown in Figure~\ref{bin8tempregs}. For the north-west front, we extracted spectra from one sector centered at 
RA=168.2268, DEC=13.4344 with PA from $7^{\circ}-122^{\circ}$ and radii $0-18.4$\arcsec\, and one with the same center 
and PA, but with radii $18.4-45.5$\arcsec, corresponding to the sector in which we fit the density model above. For the 
south-east front, we extracted spectra in one sector centered at RA=168.2197, DEC=13.45094 with PA from $240^{\circ}-267^{\circ}$ 
and radii $127.9-175.1$\arcsec, and one with the same center, but PA from $234^{\circ}-273^{\circ}$ and radii 
$175.1-270.6$\arcsec. Note that the
region from which we extracted a spectrum for outside the south-east front has a slightly larger opening angle compared to
the region in which the surface brightness profile was fitted. The sector is wider for better statistical accuracy.
Spectra and responses for these regions were extracted as described in \S\ref{global.temp}.

The spectra outside and inside the fronts were fitted successively, with the regions outside the front fitted using an 
absorbed MEKAL model with the column density and metallicity fixed to the values derived in \S\ref{global.temp}. 
The regions inside the fronts were then fitted with a model containing two absorbed MEKAL components, the primary 
MEKAL component models the gas within the front and the secondary MEKAL component accounts for gas lying in 
projection along the line of sight, which is assumed to have the same thermal properties as the gas lying in the region
outside the front. The column density and abundance were fixed to the values derived in \S 
\ref{global.temp} for both components and the secondary component had its temperature fixed to the best fit value 
obtained from the outer region spectra. The normalization of the secondary component was fixed to the value obtained 
from fitting the spectrum outside the front, corrected by a factor that accounts for the different emission measures 
expected from the different volumes probed, which is calculated by integration of the density model. For the south-east 
front, we measured a temperature of $5.7^{+3.7}_{-1.7}$\,keV outside the front and $3.6^{+1.0}_{-0.7}$\,keV inside the 
front, and for the north-west front we measured a temperature of $5.2^{+1.5}_{-1.0}$\,keV outside the front and 
$3.2^{+1.0}_{-0.7}$\,keV inside the front. 

The pressure should be continuous at a cold front. The pressure jump across a front can be measured by taking 
the ratio ($n_{e,1} kT_1)/(n_{e,2}kT_2$). We measured pressure jumps of $1.4_{-0.8}^{+1.8}$ and $1.1_{-0.6}^{+1.1}$ for the 
south-east and north-west fronts, respectively, consistent with the pressure being continuous across both fronts. A shock 
interpretation for the fronts can be also ruled out by applying the Rankine-Hugoniot shock jump conditions for
the measured density jumps and post-shock temperatures (i.e. the temperature on the dense side of the front) and comparing
with the values observed. For shock fronts, we would expect to observe pre-shock gas temperatures of  
$1.9^{+.9}_{-0.4}$\,keV and  $2.1_{-0.8}^{+1.2}$\,keV for the south-east and north-west fronts, respectively, significantly different from 
our measured temperatures. With continuous pressure and a temperature increase, these discontinuities have the hallmarks 
of cold fronts.

\subsubsection{Temperature Map}

Temperature maps provide extremely useful tools for searching for multi-phase temperature structure due to the effects of
an ongoing merger. To search for evidence of merger induced temperature structure in the ICM of Abell~1201, we generated 
a temperature map  using the broad energy band method described in \citet{markevitch2000}. Briefly, we produced source 
and background images, binned in 7.8\arcsec\, pixels, in the energy bands 0.5-1.0-2.0-5.0-10\,keV, excluding point 
sources. Exposure maps that corrected for mirror vignetting relative to the on-axis position, QEU (including low energy 
contamination) and exposure time were produced for each energy band. The background images, which were taken from the 
blank sky observations described above, were normalized by the ratio of the source to background 9-12\,keV counts, 
subtracted from the source images and the resulting image divided by its corresponding exposure map. Each image was 
smoothed using the same variable width Gaussian, where $\sigma$ varied from 8.6-39.4\arcsec, with it being smallest in 
the brightest regions and chosen such that statistically significant temperatures could be measured whilst maximum spatial
information was retained. The noise in each pixel was determined from the raw, uncorrected images and weighted accordingly
to allow for the effects of the smoothing. 

\begin{figure}
{\includegraphics[angle=-0,width=0.48\textwidth]{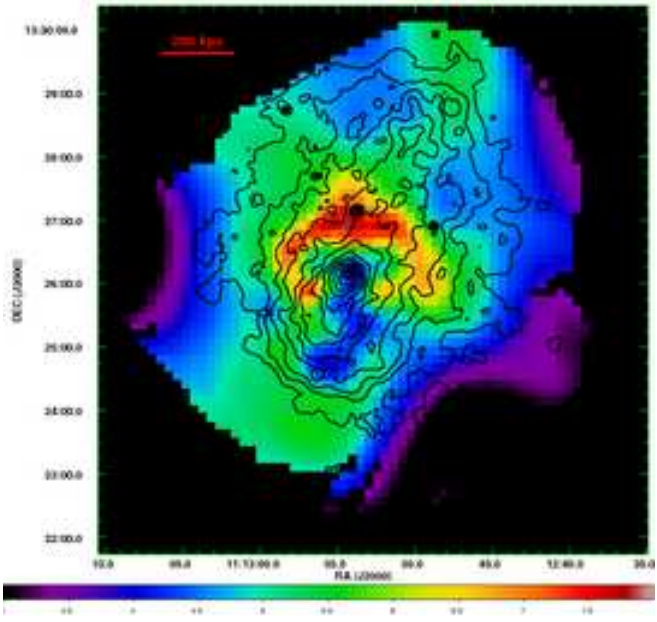}}
\caption{Temperature map with 0.5--7.0\,keV brightness contours overlaid (same as Figure~\ref{optxraycont}). The $68\%$ 
temperature uncertainties range from $\pm 0.4$\,keV in the cooler ($\rm kT<5$\,keV)  regions to 
$\pm 1-2$ in the hotter ($\rm kT>5$\,keV) regions. The color differences represent regions which are at significantly 
different temperatures. The color bar shows the temperature scale in keV.}
\label{tempmap}
\end{figure}

Each pixel was fitted with an absorbed single temperature MEKAL model with the absorption column set at the Galactic 
value. The metal abundance was set to the best fitting average cluster value derived in \S\ref{global.temp}. The 
model was multiplied by an on-axis ARF to correct for the energy dependent on-axis mirror effective area, including the 
chip quantum efficiency (QE). A flux-weighted spectral response matrix was generated from a large cluster region, and was 
binned to match the chosen energy bands. The temperature map is presented in Figure~\ref{tempmap}, where we have excluded 
pixels where the $1\sigma$ error is greater than $30\%$ of the best fit temperature value.

The cold fronts described above are clearly visible in Figure~\ref{tempmap} and the temperature map values are consistent 
with the temperatures measured spectroscopically in the above sections. Interestingly, there is a finger of cool gas 
extending from the south-east cold front to the core of the cluster. This finger is coincident with the residual observed after
subtraction of the Beta model in \S\ref{betaresid}. Also of interest is the significantly hotter $\sim 7.5$\,keV
region lying between the core and the excess subclump. The temperature in the region containing the north-west excess 
appears to be consistent with the global cluster temperature. Since the surface brightness here is low, the 
photon numbers are also low, meaning the region must be heavily smoothed to obtain statistically significant 
temperatures ($\sigma\simeq 20\arcsec$ here). Thus the temperature measurement is averaged over a large region and 
is heavily contaminated by ambient gas projected onto this region, thus the temperature of the excess remains uncertain.
 
\section{Optical Spectroscopy - Reduction and Analysis}\label{opticalanal}

In this section we present our optical analysis which includes selection of photometric samples for spectroscopic follow 
up, spectroscopic observations and data analysis, determination of cluster membership and substructure detection. The 
purpose of this optical analysis is to search for substructures and use them in correlation with the X-ray observations
to develop a scenario for the formation of the cold fronts observed above, and also for the merger history of Abell~1201.

\subsection{Data Reduction and Selection Criteria}

The multi-object spectroscopic (MOS) data presented in this paper were taken from two sets of observations, one set at 
the 3.9m Anglo-Australian Telescope (AAT) on 2006 April 2-5, and the second at the 6.5m Multiple Mirror Telescope 
(MMT) in queue schedule mode in the months of February and April 2007. The observation details are listed in 
Table~\ref{obslog} where we list the dates, magnitude limits, frame exposures and the seeing.

\begin{deluxetable*}{ccccc}
\tabletypesize{\scriptsize}
\tablecolumns{5}
\tablewidth{0pc}
\tablecaption{Summary of the observations.\label{obslog}}
\tablehead{
\colhead{Telescope/Instrument}&\colhead{Date} & \colhead{Magnitude} & \colhead{Frames} & \colhead{Seeing}}
\startdata
AAT/AAOmega&2006 Apr 2&$19.5<R<20.5$& $3 \times 1500\rm s+$& 1.5-2.0\arcsec\\ 
&&&$3 \times 750\rm s$&\\ 
\nodata&2006 Apr 4&$19.5<R<20.5$& $3 \times 1800\rm s $&2.0\arcsec \\ 
\nodata&2006 Apr 5&$R<19.5$& $3 \times 1200\rm s $ & 2.0\arcsec\\ 
\nodata&2006 Apr 5&$R<19.5$& $3 \times 1200\rm s $ &2.0\arcsec\\ 
\nodata&2006 Apr 5&$R<19.5$& $3 \times 1200\rm s $ &2.5\arcsec\\ 
MMT/Hectospec&2007 Feb 21&$20.5<R<21.5$ & $4\times 1800\rm s$& 0.6\arcsec\\
\nodata&2007 Feb 22& $20<R<20.5$&$4\times 1200\rm s$ &1.2\arcsec\\
\nodata&2007 Apr 18 &$R<20.5$ &$3\times 1200\rm s$ & 1.2\arcsec\\
\nodata&2007 Apr 19&$R<20.5$ &$4\times 1200\rm s$ & 0.8\arcsec\\
\enddata
\end{deluxetable*}

\subsubsection{Parent Photometric Catalog}

The initial target catalog was taken from the Sloan Digital Sky Survey (SDSS) sky 
server\footnote{See: http://cas.sdss.org/dr5/en/}, with all objects within an 18\,arcmin radius of the center of 
Abell~1201 (RA=$11^{\rm h}12^{\rm m}54.5^{\rm s}$, DEC=+$13^{\circ}$$26{\arcmin}$$09.0{\arcsec}$) included. This catalog was then filtered to include only those objects 
classified by the SDSS pipeline as galaxies (SDSS class 3).  The SDSS $r$ and $g$ magnitudes for the remaining objects 
were then converted to Johnson $B$ and $R$ magnitudes via equations A5 and A7 of \citet{cross2004} and only those galaxies
with R $< 21.5$ were retained in the final target list. 

The target galaxies were then ranked based on their cluster-centric radius and their position on the B-R vs R 
color-magnitude (CM) diagram, in preparation for the fiber allocation procedure. Galaxies on or blueward of the red 
sequence are more likely to be cluster members and so were ranked higher than those redward of the red sequence. 
Those galaxies either on or blueward of the red sequence were then ranked by cluster-centric distance, with those 
closest to the center ranked highest. The remaining galaxies lying redward of the red sequence were also ranked by 
cluster-centric distance. These rankings were input into the software used to configure the fiber allocations for the 
respective observations, to ensure that higher weightings were placed on galaxies which were more likely to be cluster 
members (i.e. those on and blueward of the red sequence and closer to the center). We discuss the configuration process 
further below. 

\begin{figure}
{\includegraphics[angle=-90,width=0.48\textwidth]{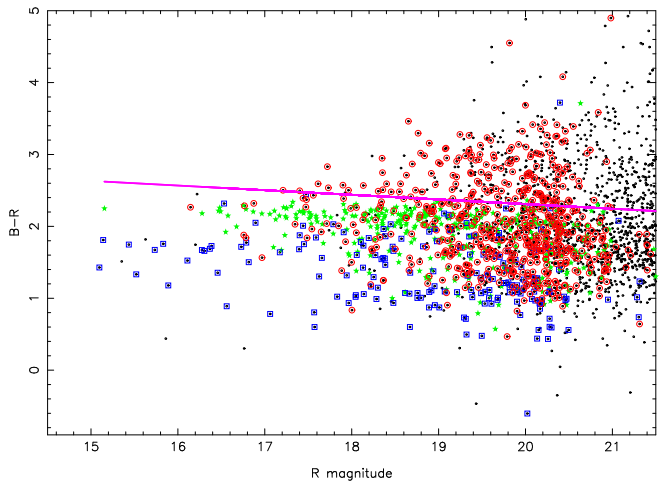}}
\caption{Color-magnitude diagram for all galaxies within our observed field (black dots). The {\it green stars} represent 
cluster members, the {\it blue squares} foreground galaxies, and the {\it red circles} background galaxies. The {\it pink 
line} is the line used to delineate galaxies on and blueward of the red sequence from those redward of the red sequence 
for the purpose of ranking galaxies during fiber configuration.}
\label{cm}
\end{figure}

We show the CM diagram in Figure~\ref{cm}, where we have overplotted the spectroscopically confirmed cluster members 
({\it green stars}), foreground galaxies ({\it blue squares}) and background galaxies ({\it red circles}). Also plotted is 
the line $B-R=3.6-0.0645R$ which was used to distinguish those galaxies which we define to be lying on or below the red 
sequence from those redward of the red sequence. The slope of the line is calculated from the best fit to the CM slope 
versus redshift diagram of \citet{lopezcruz2004} and the constant is estimated by eye such that the line lies just above 
the red sequence.

\subsubsection{AAT AAOmega Observations}

The AAT observations were taken using the AAOmega fiber-fed spectrograph \citep{saunders2004,smith2004,sharp2006}, 
which is a bench mounted dual-beam spectrograph, which is fed by 400 fibers robotically-placed within the two degree field
at the telescope's prime focus. A total of 392 fibers, each 2\arcsec\, in diameter, are available for the simultaneous 
observation of scientific targets (with the remaining 8 fibers being used for acquisition and guiding). All our 
observations were taken using the medium resolution ($R\simeq1300$) 580V (blue arm) and 385R (red arm) gratings, which in 
combination with the $15\,\mu$m pixel  2k$\times$4k E2V CCD detectors produce a spectral resolution of $3.6$\AA\, and 
$5.5$\AA\, in the blue and red arms, respectively, and provide an overall wavelength coverage ranging from 3700 to 8800\AA.

Due to the clustered nature of the galaxies, fiber positioning constraints (minimum separation of $\sim30$\arcsec\, due to 
the physical size of the fiber buttons) and the fact our input catalog contained $\sim1400$ galaxies, multiple plate 
configurations were required to obtain high overall spectral completeness and adequate coverage in the central regions of 
the cluster. The fibers were allocated using the AAOmega {\it CONFIGURE} 
software\footnote{See: http://www.aao.gov.au/AAO/2df/}. A total of 5 configurations were 
required for Abell~1201: three for galaxies with $R < 19.5$, and two for galaxies in the range $19.5 < R < 20.5$  
plus those that remained unallocated in the three `bright' configurations. Due to packing constraints, only  110-210 of 
the available fibers were allocated to science objects. In addition, 30-40 fibers were allocated to blank sky regions for 
sky subtraction. All fibers known to suffer from interference fringing effects were left unallocated. The details for each
configuration observation are listed in Table~\ref{obslog}. The observations were taken in rather mediocre conditions, 
with the seeing (FWHM) ranging from 1.5 to 2.5\,arcsec, and all exposures being taken through thin cloud cover, with some 
being interrupted by thicker clouds. In commencing the observations for each new configuration,  a 4\,s dome flat and 
60\,s FeAr arc lamp exposure were taken for flat fielding and wavelength calibration of the data, respectively. The data 
were reduced with the AAO {\it 2dFDR} pipeline software with a patch included to fix fiber to slit position mapping errors
present in AAOmega data taken before August 2006. 

Redshift identification and measurement for each spectrum was carried out using the RUNZ code written by Will Sutherland 
for the 2dF Galaxy Redshift Survey \citep[2dFGRS;][]{colless2001}. This program utilizes the cross-correlation method of 
\citet{tonry1979}, based on a library of galaxy template spectra that are representative of all the different observed 
spectral types. Each spectrum was inspected visually and given a redshift quality classification, Q. Here we used the 
same scheme as adopted for the 2dFGRS, with each spectrum being assigned a Q value on a six-point integer scale, with 
Q=1 indicating that no redshift could be estimated, Q=2 a possible but unreliable redshift, Q=3 a probable redshift 
(with $\sim 90\%$ confidence), Q=4 a reliable redshift, Q=5 a reliable redshift with high-quality spectrum, and Q=6 
indicating a star or non-extragalactic object. We obtained spectra for 917 galaxies during the run, which yielded 
reliable (Q=3, 4, or 5) redshift measurements for 580 single galaxies.

\subsubsection{MMT Hectospec Observations}

The MMT observations were taken using the Hectospec multi-object spectrograph \citep{fabricant2005}. This is also
a bench-mounted spectrograph, which is fed by 300 1.5\arcsec\, diameter fibers that cover a 1\,degree field of view.  
The observations were taken using the medium resolution (R=1000-2000) 270 groove ${\rm mm}^{-1}$ grating and the 
data were captured on a single array of two E2V CCDs with 13.5$\mu m$ pixels, resulting in spectra with a resolution of 
6.2\AA\,  and covering the wavelength range $3500-10000$\AA.

For the MMT observations, we included any galaxy not observed during the AAT run, as well as galaxies that were not 
assigned reliable redshifts. A small number of  galaxies which were assigned a redshift quality of  Q=3 (from their 
AAT spectra) and were in the redshift range 0.15-0.18 (i.e. close to the cluster in redshift space), were also included in 
order to check on our redshift accuracy. Fiber configurations were generated using the XFITFIBS 
software\footnote{see:http://cfa-www.harvard.edu/mmti/hectospec.html}, with a total of four being required for galaxies 
with $R < 20.5$, and one being required for galaxies with $20.5<R<21.5$. Typically 150 fibers were allocated to galaxies, 
with around 100 fibers allocated to blank sky areas. The details of each observed configuration are listed in 
Table~\ref{obslog}. The seeing during the observations ranged from 0.61-1.2\arcsec, and the 
observations performed in the worst seeing conditions were somewhat affected with a number of galaxies requiring 
re-observation since the spectra were inadequate for determining redshifts. The data were reduced at the Telescope Data 
Center (TDC)\footnote{see:http://tdc-www.harvard.edu/} using 
the TDC pipeline. The data were also redshifted at the TDC using the IRAF cross-correlation XCSAO software 
\citep{kurtz1992} and spectra were assigned a redshift quality of ``Q'' for reliable, ``?'' for questionable and ``X'' 
for bad redshift measurements. We use in our analysis only those galaxies with reliable redshift measurements (quality=Q).
A subsample of the data was visually inspected and the redshifts and quality assignments were found to be robust. We 
obtained spectra for 742 galaxies during the run, which yielded reliable redshift measurements for 534 single galaxies.

\subsubsection{Redshift completeness and measurement errors }

A number of galaxies were observed multiple times in the two observing runs. There were also 67 galaxies
with independently measured redshifts within 20\,arcmin of the cluster center in the NASA Extragalactic Database 
(NED)\footnote{see:http://nedwww.ipac.caltech.edu/}. To check the accuracy of our measured redshifts, we compared the
redshifts of all multiply-observed galaxies by determining the mean of the differences as well as their standard deviation.
The MMT observations had a mean intrinsic uncertainty of $42\pm1$\kms\, for 534 redshifts. There were 19 repeat 
observations made with the MMT, and the mean difference was $1\pm25$\kms\, after removing a spurious result where the 
difference was $cz=44894$\kms. From these repeat observations, we calculate an RMS of 105\kms\, implying an uncertainty 
of 74\kms\, for the MMT observations, higher than the mean of the individual redshift measurements. An external test of 
these redshift uncertainty measurements comes from the 12 redshifts in common between the NED and MMT catalogs which have 
a mean difference of $-28\pm20$\kms\, with an RMS scatter of 74\kms. The NED redshifts come primarily from the SDSS and 
the catalog of \citet{miller2002} which have redshift uncertainties $\sim30$\kms. Taking this into account gives an 
external uncertainty measurement of 43\kms, consistent with the mean of the individual uncertainty measurements. The high 
uncertainty value given by the repeat MMT observations can be understood by considering that these objects were 
re-observed due to the low quality of their initial redshift measurements. 

There were no repeat observations taken at the AAT, which have a mean internal uncertainty for the individual redshift 
measurements of $109\pm6$\kms\, from 580 redshifts. An external check of this value comes from 31 galaxies re-observed 
with the MMT which had both Q=3 redshifts from the AAT data set and quality=Q from the MMT. The mean difference in the 
measured redshifts of these was $50\pm44$\kms. Two galaxies with obviously spurious redshift differences of 
$\Delta(cz)=53417$\kms\, and $\Delta(cz)=119800$\kms\,were excluded from this calculation. The RMS scatter in the 
difference was 248\kms, and taking into account the uncertainty derived above for the MMT redshifts (74\kms), we calculate
an uncertainty of 159\kms. This value is higher than the mean of the individual redshift uncertainties, however again it 
is noted that these objects were re-observed due to their poor initial redshift measurements (all had Q=3), thus the 
measured uncertainty is expected to be higher. This value can be checked using the 45 measurements in common with the NED 
catalog, where the mean difference was $33\pm29$\kms, and an RMS scatter in the differences of 191\kms, after two spurious
results where the redshift differences were $\Delta(cz)=3480$\kms\,and $\Delta(cz)=10472$\kms\,were excluded. Based on 
these redshift differences, we measure an uncertainty of 132\kms, again after accounting for the $\sim30$\kms\, uncertainty
associated with the NED redshifts. 

\begin{figure}
{\includegraphics[angle=-90,width=0.48\textwidth]{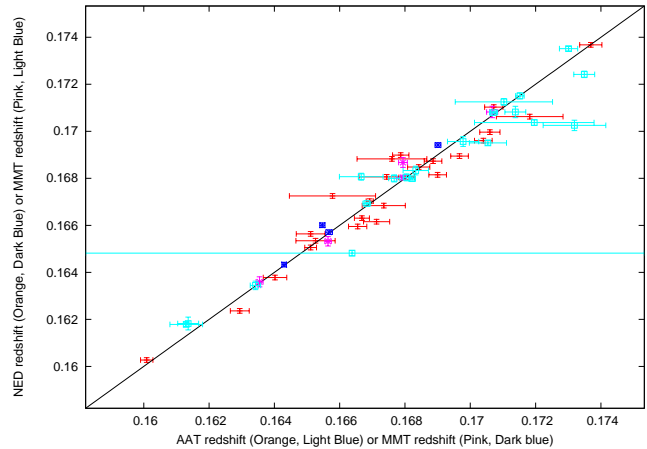}}
\caption{Redshift comparison for Abell~1201 using our MMT and AAT redshifts, and also a sample of NED redshifts. The 
comparison of different measurements are plotted with the following color codes: MMT/MMT = pink, AAT/MMT = light blue, 
AAT/NED = orange, MMT/NED = dark blue. The black line represents the one-to-one relationship; the general scatter of the 
data about this line indicates the absence of any gross systematic errors in our measurements, and indicates our AAT and 
MMT measurements have a precision of  159\kms\, and 74\kms, respectively. We assume an uncertainty of 30\kms\,for the 
NED measurements.}
\label{zcomp}
\end{figure}

Figure~\ref{zcomp} shows the comparison of redshift measurements using different instruments in the 
redshift range of the cluster. The figure, in combination with the redshift differences above, show that there are no 
gross systematic errors in our measurements and the scatter about a 1-to-1 relationship is well encompassed by the 
uncertainties of 159\kms and 74\kms for AAT and MMT, respectively, which we take from the more conservative higher values 
derived above from the double observations, and these values indicate the precision of our measurements.

For galaxies which had multiple redshift measurements, the following approach was taken to determining what
their final adopted redshift would be: If a galaxy was observed with both the MMT and AAT, we used the MMT 
redshift since the MMT redshifts were in general more precise with lower errors (as expected since the re-observed 
galaxies were ones with AAT quality Q=3). If a galaxy was observed more than once with the MMT, we took the redshift based 
on the highest cross-correlation coefficient. If a galaxy was observed more than once with the AAT, we took the redshift 
with the highest Q value. Note that in no cases did we ever use the NED redshift if that galaxy had been observed on 
either the AAT or MMT. When added to our single redshift measurements, this yielded a total of 560, 534 and 10 redshifts 
measured with the AAT, MMT and sourced from NED, respectively, giving an overall total of 1104 robust redshifts acquired 
within the cluster field.

\begin{figure*}
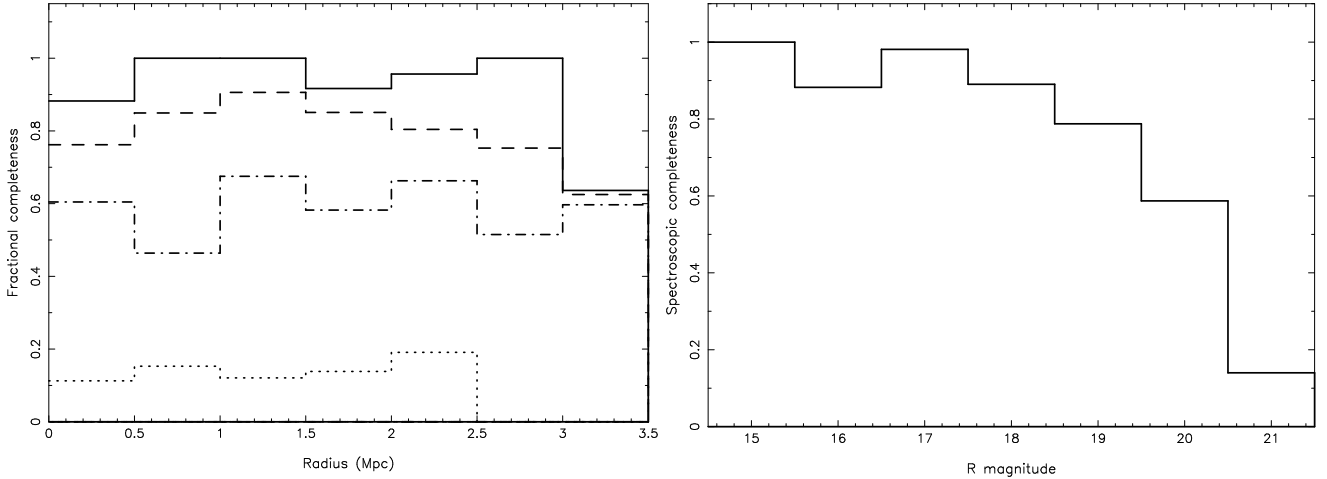

  \begin{tabular}{cc}
    {\includegraphics[angle=-90,width=0.48\textwidth]{compl.vs.rad.ps}}
    {\includegraphics[angle=-90,width=0.48\textwidth]{compl.vs.rmag.ps}}\\
  \end{tabular}
  \caption{Spectroscopic completeness as a function of cluster-centric radius ({\it left} panel) for the $R$-band magnitude intervals
0-18 ({\it solid} line), 18-19.5 ({\it dashed} line), 19.5-20.5 ({\it dot-dashed} line), and 20.5-21.5 ({\it dotted} line). This shows we 
have both good completeness and radial coverage for magnitudes brighter than $R=20.5$. The spectroscopic completeness as a 
function of magnitude is shown in the {\it right} panel.}
\label{compl}
\end{figure*}

We determined the spectroscopic completeness of the sample by measuring the ratio of galaxies in the parent photometric 
catalog which have reliable redshifts (as defined previously) to those which have no reliable redshift measurement. The 
spectroscopic completeness as a function of cluster-centric radius for the $R$ magnitude intervals 0-18, 18-19.5, 19.5-20.5, 
and 20.5-21.5 is plotted in Figure~\ref{compl} where we also plot the spectroscopic completeness within 3.5\,Mpc as a function 
of $R$ magnitude. We achieve $\gtrsim 80\%$ spectroscopic completeness at all radii, apart from within the 3-3.5\,Mpc range, 
for magnitudes brighter than 19.5, whilst we obtain $\sim 60\%$ spectroscopic completeness at all radii for magnitudes 
between 19.5 and 20.5 and $\sim 20\%$ spectroscopic completeness for magnitudes between 20.5 and 21.5 within a 
radius of $\sim 2.5$\,Mpc. Thus our catalog is very well sampled at all radii for $R$ magnitudes brighter than 20.5, 
which corresponds to $\sim 2.3$\,mags down the cluster luminosity function (assuming $M^*_R = -21.3$ \citep{yagi2002}, 
and neglecting galactic extinction and K-correction terms). 

\subsection{Cluster member selection}\label{membersel}

\begin{figure}
{\includegraphics[angle=-90,width=0.48\textwidth]{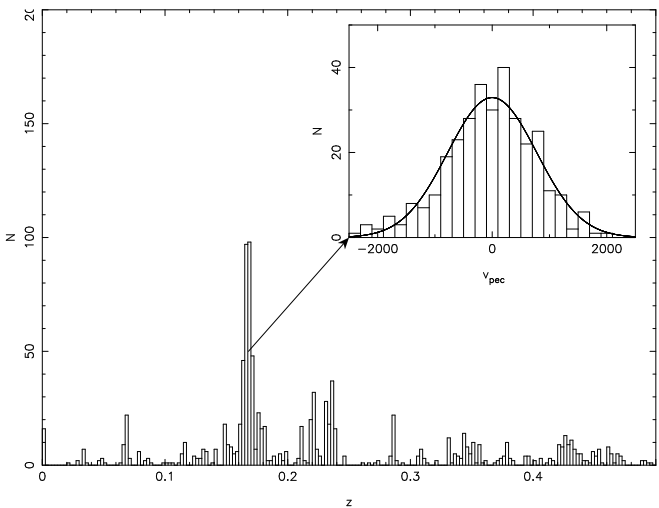}}
\caption{A histogram of all the reliable redshifts measured within the Abell~1201 field. The Abell~1201 cluster 
clearly stands out against the foreground and background galaxies. The inset panel shows the distribution of 
peculiar velocities for the cluster members, with a Gaussian with mean zero and standard deviation of 778\kms\, 
overplotted.}
\label{allhisto}
\end{figure}

Identification of cluster members from our spectroscopic redshifts was achieved through the elimination of
foreground and background galaxies along the line of sight to Abell~1201. An initial rejection was performed
using the ``velocity gap'' method outlined by \citet{depropris2002}, where the galaxies are sorted in redshift space 
and the velocity ($cz$) gap between each one determined. Here the velocity gap for the nth galaxy is 
$\Delta v_n = cz_{n+1}-cz_n$. Clusters appear as well populated peaks in redshift space which are separated by velocity 
gaps of greater than 1000\kms\,from the nearest foreground and background galaxies, and Figure~\ref{allhisto} shows that 
Abell~1201 is readily identified on this basis. 

Due to the filamentary structures which surround clusters, it is necessary to further refine this rejection 
process. For this refinement, we used a slightly different version of the ``shifting gapper'' method first implemented 
by  \citet{fadda1996} where both cluster-centric radius and peculiar velocity information are used and the gap method 
outlined above is applied as a function of radius. The galaxies were binned radially such that each bin contained 35 
galaxies, and were then sorted by $v_{pec}$, with the velocity gaps determined as before (but in peculiar velocity, not 
$cz$). Peculiar velocities were determined with respect to the central cluster redshift using the following procedure: We 
estimated the mean cluster redshift using the biweight location estimator \citep{beers1990} which we assume represents 
the cosmological redshift of the cluster. The peculiar redshift, $z_{pec}$, was derived under the assumption that the 
observed redshift of the galaxy, $z_{gal}$, is comprised of only two components, the cosmological component, $z_{cos}$, 
and the component due to the peculiar motion within the cluster, $z_{pec}$. Hence, the peculiar redshift is 
$z_{pec} = (z_{gal} - z_{cos})/(1+z_{cos})$ and the peculiar velocity was derived using the standard special relativistic 
formula $v_{pec} = c((1+z_{pec})^2-1)/((1+z_{pec})^2+1)$, where $c$ is the speed of light.

\begin{figure}
{\includegraphics[angle=-90,width=0.48\textwidth]{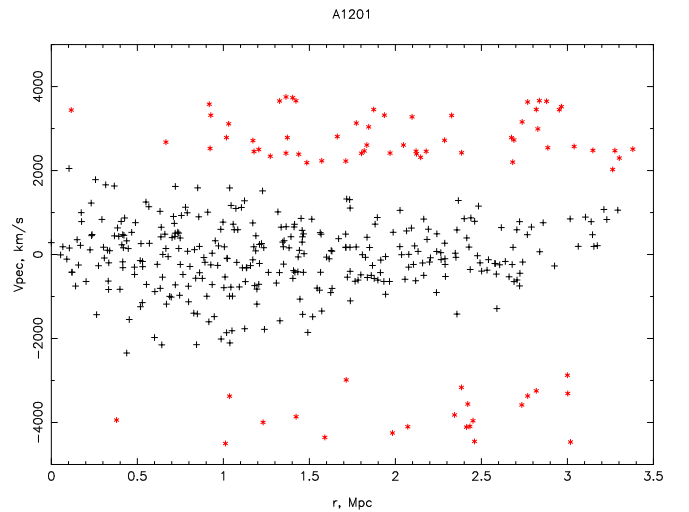}}
\caption{A plot of deviations in peculiar velocities as a function of cluster-centric radius, illustrating the
efficacy of our refined  rejection method whereby foreground and background interlopers are identified
and eliminated using a shifting gapper technique (see text). The {\it black} crosses represent galaxies 
allocated as cluster members while the {\it orange} asterisks are rejected foreground and background
galaxies lying close to the cluster in redshift space.}
\label{shiftgap}
\end{figure}

We used the ``f pseudosigma'' \citep{beers1990}, derived from the first and third quartiles of the peculiar velocity 
distribution, as the fixed gap to separate the cluster from interlopers. The f pseudo-sigma is an estimator of the scale of 
the velocity distribution which is robust to the presence of velocity interlopers in the tails of the cluster distribution. 
The above procedure was iterated until the number of members was stable and the results are shown in Figure~\ref{shiftgap} 
where it can be seen that the cluster is clearly separated from the filamentary structure surrounding it, and the 
interlopers are cleanly rejected. The advantage of this method is that it does not assume a particular mass model or 
velocity distribution which both rely on the cluster being relaxed.

The final cluster sample contains 321 members out to a cluster-centric radius of $\sim3.5$\,Mpc. The final value for the 
biweight location estimator of the cluster redshift is $z_{cos}=0.1673\pm0.0002$. We use the biweight scale estimator 
\citep{beers1990} to estimate a velocity dispersion of $778\pm36$\kms. The errors for the redshift and velocity 
dispersion are $1\sigma$ and are estimated using the jackknife resampling technique.

\subsection{Substructure Detection}

Now that we have separated our cluster members from foreground and background interlopers, we can apply substructure
detection tests to the cluster member sample. Substructure in a cluster can present itself in a number of different ways, 
and generally no single statistical test is capable of revealing all such manifestations. Hence it is essential that the 
full range of statistical
tests are applied if the search for substructure is to be an exhaustive one. In general, statistics using the 
maximum amount of information (eg., the radial velocity and spatial dimensions) are the most effective 
at revealing substructure, however these can fail if, for example, roughly equal 
mass clusters are merging along the line of sight such that the cores are spatially coincident 
\citep[see eg.][]{girardi2006,pinkney1996}. The use of velocity information by itself can be successful, although there are 
documented cases where the velocity distribution mimics that of a relaxed cluster where other methods 
have clearly shown it to be disturbed \citep{maurogordato2000,johnstonhollit2008}. The use of 2-D spatial information 
alone suffers from foreground and background contamination. In this section, we apply a number of statistical methods 
to the detection of substructure in Abell~1201. 

\subsubsection{Deviations from Gaussianity}

The velocity distribution of a relaxed cluster is well approximated by a single Gaussian, with velocity dispersion 
related to the cluster mass via the virial theorem. Departures from 
Gaussianity can be attributed to a dynamically active cluster where substructures may cause symmetric and asymmetric 
distortions in the velocity distribution. As a first test of Gaussianity, we used the standard Kolmogorov-Smirnov (K-S) 
test.  At the $90\%$ confidence level, the observed velocity distribution is consistent with being drawn from a Gaussian 
distribution with $\overline{v_{pec}}=0$ and $\sigma_{v{pec}}=778$\kms. On the basis of this test, therefore, it appears 
unlikely Abell~1201's velocity distribution is significantly non-Gaussian.

The disadvantage of using the K-S test is that it is most sensitive to the behavior of the distribution
near its median, but is relatively insensitive to differences in the tails of the distribution. Also, the K-S test does 
not give quantitative information about the way in which two distributions differ. Quantifying these deviations is 
imperative in determining the state of the cluster, and a number of different methods can be used to quantify asymmetric 
and symmetric distortions \citep{pinkney1996}. Here we choose to use the method outlined in \citet{zabludoff1993}, 
where the velocity distribution, $L$, is approximated by a series of Gauss-Hermite functions
\begin{equation}
L=\sum^{4}_{j=0} h_j H_j(x),
\label{veldist}
\end{equation}
where $h_j$ are the Gauss-Hermite moments defined by \citet{zabludoff1993} as
\begin{equation}
h_j = \frac{2 \sqrt{\pi}}{NS} \sum^{N}_{i=1} H_j(x_i),
\label{hermites}
\end{equation} 
where
\begin{equation}
x_i = \frac{{v}_{pec,i}-V}{S},\, H_j(x)=\frac{e^{-x^2/2}}{\sqrt{2\pi}}\mathcal H_j(x),
\end{equation} 
and $\mathcal H_j(x)$ are the Hermite polynomials given by \citet{vandermarel1993}. In principle, the velocity 
distribution can be described by equations \ref{veldist} and \ref{hermites} and a wide variety of $S$ and $V$ values, 
where there is significant degeneracy in the choices of $S$ and $V$ and the $j=1,2$ terms. However, as 
pointed out by \citet{vandermarel1993}, the most efficient method is to set $S$ and $V$ so that the zeroth-order term, 
$H_0$, describes the best fit Gaussian to the data. This is achieved by choosing $S$ and $V$ such that $h_1=h_2=0$. We 
iterated with different values of $S$ and $V$ until these criteria were met. The terms $h_3$ and $h_4$ describe, 
respectively, the asymmetric and symmetric deviations from Gaussianity, much like the higher order Gaussian skewness and 
kurtosis terms, but have the advantage of being less sensitive to outliers in the tails of the distribution. Applying 
this method to the velocity distribution for Abell~1201, we obtain values of values of $h_3=-0.016$ and $h_4=0.023$. We 
used 10,000 Monte Carlo realizations of Gaussian distributions with N=321, $\sigma=S$ and $\mu=V$ to determine the 
probability of detecting our  $h_3$ and $h_4$ terms. Values of $|h_3|\ge0.016$ occur in $70\%$ of the realizations, while
values of $|h_4|\ge0.023$ occur in $50\%$ of the realizations. We conclude there is no significant evidence for asymmetric
or symmetric deviations from Gaussianity.

\subsubsection{Velocity Distribution Profiles}

\begin{figure}
{\includegraphics[angle=-90,width=0.48\textwidth]{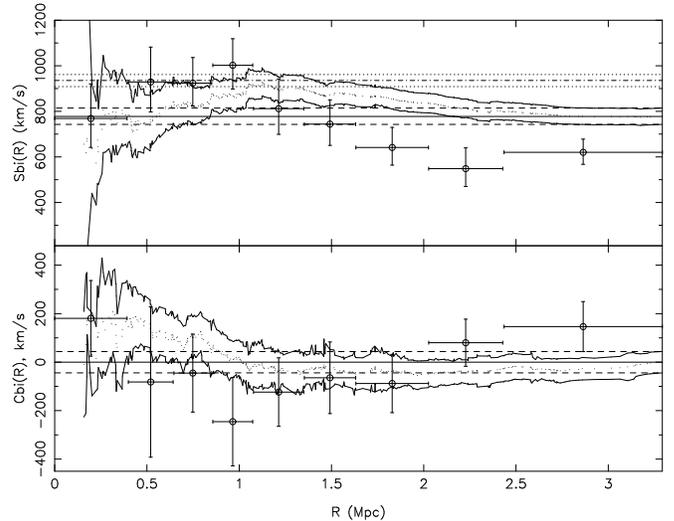}}
\caption{Differential velocity dispersion profile ({\it top} panel) and mean peculiar velocity ({\it bottom} panel) profiles
as a function of projected radius. The {\it solid} lines show the total cluster values, while the {\it dashed} lines and error bars show 
the $1\sigma$ confidence limits derived using the jackknife technique. The {\it dot-dashed} line shows the velocity dispersion 
derived from the mean X-ray temperature (see text), with the {\it dotted} lines showing the upper and lower limits based on the 
errors in the temperature (see \S~\ref{global.temp})}
\label{vprofs}
\end{figure}

Figure~\ref{vprofs} shows integral and differential projected radial profiles of the  biweight location, $\mu(R)$ and 
scale, $\sigma(R)$, estimators. The differential profiles are binned using the same radial intervals as used for the shift 
gapper in \S\ref{membersel}, such that each bin contains at least 35 data points. To measure the integrated profiles, we 
sorted the data by cluster-centric radius and for each galaxy in the sample, starting from the tenth, we measured the 
biweight location and scale (where the number of galaxies was less than 15, we used the median and gapper estimates of 
location and scale) using only data within the radius of interest to each galaxy.

The integrated and differential $\mu(R)$ are constant with radius within the $1\sigma$ error bars. This
is expected for an isotropic distribution of radial velocities. The differential $\sigma(R)$ profile is flat out to 
$r=1$\,Mpc, and smoothly declines at larger radii. \citet{denhartog1996} classified $\sigma(R)$ profiles as flat, inverted or
peaked, depending on the shape of the profile within $\sim 1$\,Mpc, and they note an interesting subset of clusters with
flat $\sigma(R)$ profiles which have the highest $\sigma(R)$ value measured between 0.5 and 1.0\,Mpc, all of which show
signs of recent merger activity. While this is by no means a definitive sign of merger activity, it is interesting 
to note that Abell~1201 also exhibits its highest value of $\sigma(R)$ at $\sim1$\,Mpc. The smooth decline outside of
1\,Mpc is expected since the escape velocity of a cluster declines with radius, while radial orbits of infalling galaxies
enhance the decline. The integral $\sigma(R)$ profile increases gradually to its maximum value at $\sim1$\,Mpc, then 
steadily decreases before flattening out at a radius $\sim2.5$\,Mpc, indicating the $\sigma$ measured at this radius is
no longer affected by velocity anisotropy, and is representative of the cluster potential \citep{fadda1996}.

We include the estimate of the velocity dispersion derived from the cluster X-ray temperature measured in \S 
\ref{global.temp} by assuming $\beta=\mu m_p\sigma^2/kT=1$ (i.e., assuming the same specific kinetic energy in the gas 
and galaxies), where $\sigma$ is the galaxy velocity dispersion, $\mu$ is the mean molecular weight of the gas particles, 
$m_p$ is the proton mass and $kT$ is the gas temperature within $R\simeq 500$\,kpc. The 
integral and differential $\sigma(R)$ profiles appear, to within the errors,  to be consistent with the value estimated from 
the X-ray temperature when considering regions within 1\,Mpc.

\subsubsection{3-D tests for sub-structure}\label{3dstructures}

The most efficient way to detect real physical substructures is to use a combination of the 2-D spatial and 1-D velocity 
information to search for local variations in the velocity distribution \citep{pinkney1996}. The most common method 
here is to utilize the $\Delta$ statistic developed by \citet{dressler1988}, which tests for differences in the 
local mean and dispersion compared to the global mean and dispersion. The downfall of this method is that it assumes the 
global and local peculiar velocity distributions are Gaussian, which may not be true for dynamically active systems. Here 
we prefer the k-statistic, $\kappa$, employed by \citet{colless1996}, which is very similar to the $\Delta$ statistic 
but does not  require the assumption of Gaussianity to be made. To determine $\kappa$, the $n=\sqrt{N}$ nearest neighbors
in projection are selected for each cluster member, where $N$ is the total cluster member sample size, and the local 
velocity distribution of the $n$ nearest neighbors is compared to the global cluster velocity distribution (minus the 
$n$ nearest-neighbor velocities). Local departures from the global velocity distribution are quantified using the K-S D 
statistic, with the null hypothesis that the local distribution is drawn from the global one and the significance 
determined by measuring the probability that the D statistic is larger than the observed D statistic for the observed 
sample size, $P_{KS}(D>D_{obs})$. Then, $\kappa_n$ is defined as
\begin{equation}
\kappa_n=\sum^N_{i=1}-{\rm log} P_{KS}(D>D_{obs}),
\end{equation}
giving a global measure of the substructure present in the cluster by summing the individual $\kappa$ values. The 
significance of $\kappa_n$ was determined by performing 10,000 Monte Carlo realizations with the peculiar velocities 
randomly shuffled, whilst maintaining the positional information, and remeasuring $\kappa_n$. 

\begin{figure}
{\includegraphics[angle=-90,width=0.48\textwidth]{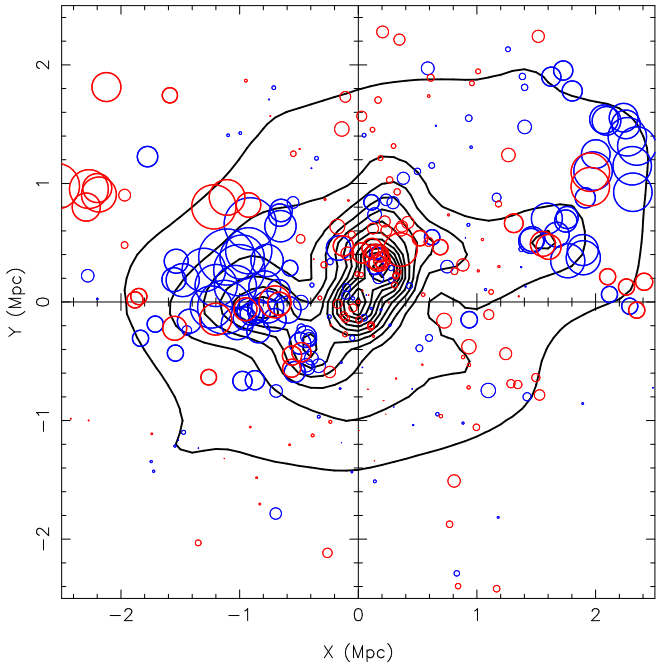}}
\caption{Bubble plot outputs from the $\kappa$ test. The {\it bold} bubbles are those deemed to be significant
insomuch as they only occur in $5\%$ of 10,000 realizations. The {\it blue} bubbles are those galaxies which have negative
$v_{pec}$, and the {\it red} bubbles have positive $v_{pec}$. The contours are galaxy density contours generated from 
applying a variable width Gaussian filter to the spatial distribution of spectroscopically confirmed members. The contours 
are linearly spaced by 10 in the interval 10-150 $\rm{gals\,Mpc}^{-2}$. Note the four clumpings of significant bubbles are
coincident with overdensities in the projected galaxy density. The center of the cluster is located at 0,0 Mpc.}
\label{bubbleplot}
\end{figure}

The observed value of $\kappa_n=379$, is larger than any value obtained in our 10,000 realizations, which follow a 
log-normal distribution with a mean value of $\mu(\ln \kappa_{sim})=4.9$ and 
standard deviation $\sigma(\ln \kappa_{sim})=0.2$. Thus, the observed $\kappa_n$ lies 
$6\sigma$ from the mean of the realizations, and we can put an upper limit on the probability of observing this $\kappa_n$ value by
chance at less than $10^{-4}$. {\it We conclude, therefore, that there is velocity substructure present in Abell~1201 at 
high significance.} The results of the $\kappa$ test are best presented using ``bubble plots'' showing a circle with 
radius $r \propto -\log P_{KS}(D>D_{obs})$ at each galaxy position, such that clustered large bubbles reveal local 
departures from the global velocity distribution. We show these bubbles in Figure~\ref{bubbleplot} and color code them 
based on the sign of the peculiar velocity, with blue and red having negative and positive $v_{pec}$, respectively. 
Overplotted are contours of galaxy surface density which have been produced by applying a variable width Gaussian filter, 
with $\sigma$ varying from $\sim 100$\,kpc in the cluster center to $\sim 400$\,kpc in the outskirts, to the spatial 
distribution of the spectroscopically confirmed members. We define significant values of $-\log P_{KS}(D>D_{obs})$ as those
which occur only $5\%$ of the time in the 10,000 Monte Carlo realizations, and these are highlighted by the bold bubbles 
in Figure~\ref{bubbleplot}.
Visual inspection of the distribution of large clustered bubbles in Figure~\ref{bubbleplot} confirms the 
significance of the measured $\kappa_n$, as there appear to be four conglomerations of significant bubbles, each 
coincident with an increase in projected galaxy density. 

Having verified the existence of substructure within Abell~1201, we further utilize the spatial and velocity information 
of our cluster galaxies by allocating group membership using the Kaye's Mixture Model (KMM) algorithm of 
\citet{ashman1994}. The algorithm fits a user specified number of N-dimensional Gaussians to the data and determines the 
improvement of the fit over that of a single N-dimensional Gaussian via a maximum likelihood test. The major drawbacks of 
this method are that the spatial distribution of the galaxies does not follow a Gaussian shape (although the velocity 
distribution does for a relaxed cluster), and the number of Gaussians needs to be known a priori. Visual 
inspection of the projected X and Y galaxy distributions reveal they are at least qualitatively Gaussian, and given the 
benefit of including an extra two dimensions in the analysis far outweighs the false assumption of Gaussianity, we 
proceed with the full 3-D KMM analysis. Overcoming the latter drawback requires a robust method of estimating the initial 
number of Gaussians, and also their parameters. Given the correlation between the significant bubbles and galaxy surface 
density peaks seen in Figure~\ref{bubbleplot}, we use it as a guide to estimate the positions and projected radii of 
substructures. There appear to be 6 spatially separated substructures where the local velocity distribution differs 
significantly from the global one, along with the main Abell~1201 cluster. We inspect the velocity distributions of all 
galaxies within the estimated projected radius for each substructure, excise any obvious interlopers and determine the 
median and standard deviation of the X position (kpc), Y position (kpc) and velocity distributions for the remaining 
substructure galaxies. These parameters serve as initial estimates for input into the KMM algorithm, and are presented in 
table \ref{kmmtable}, along with the outputs from the KMM algorithm, where ($\overline{x},\overline{y}, \overline{v}$) 
are the means of the distributions, ($\sigma_{x}, \sigma_{y}, \sigma_{v}$) the standard deviations, $N_{gal}$ is the 
number of galaxies in the substructure and Rate is the estimate of the overall rate for correct allocation of galaxies to 
this substructure. 

\begin{figure}
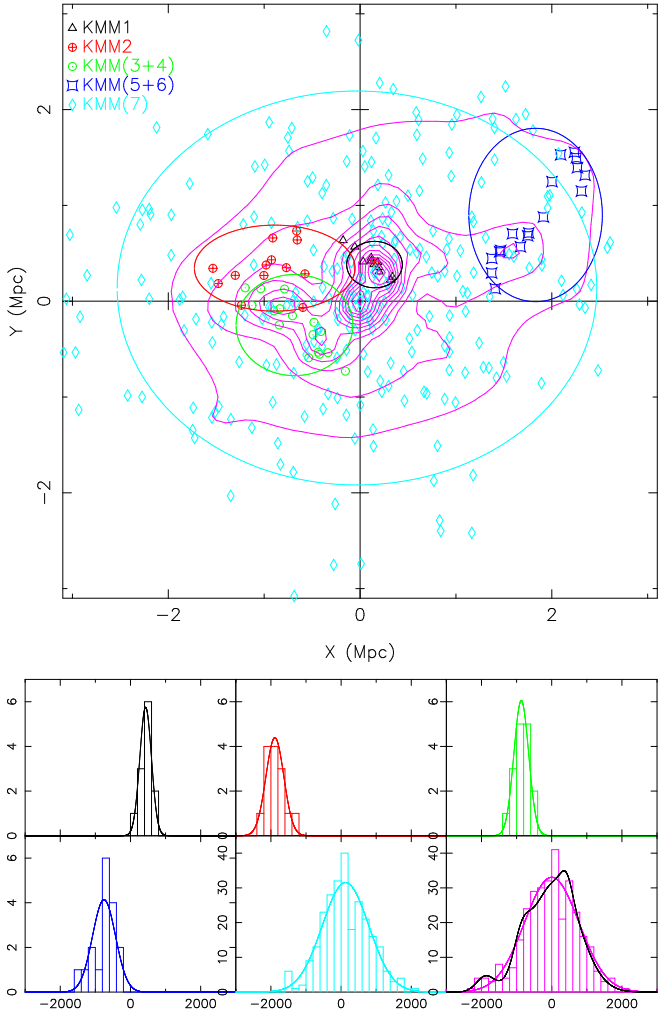

\begin{tabular}{c}
{\includegraphics[angle=-90,width=.48\textwidth]{kmm5.densconts.ps}}\\
{\includegraphics[angle=-90,width=.48\textwidth]{crap.mpc5.ps}}
\end{tabular}
\caption{{\it Top Panel:} The spatial distribution of the different partitions assigned by the KMM analysis. The 
ellipses show the 2$\sigma$ contours for the Gaussians fitted to the respective spatial distributions. We also plot the 
galaxy surface density contours in {\it pink} (same spacing as Fig. \ref{bubbleplot}). {\it Bottom panel:} The velocity 
distributions of the different partitions; the {\it curved lines} represent Gaussian functions whose mean and
$\sigma$ are equal to the KMM $\overline{v}$ and $\sigma_v$ values. 
The color coding in the bottom panel matches the key in the top left of the top panel. The bottom 
right-most velocity distribution shows the whole cluster sample with a Gaussian generated using the biweight estimators 
overplotted in {\it pink}; the combination of all the KMM Gaussians is plotted in {\it black}.}
\label{kmmplot}
\end{figure}

Given that the substructures KMM3 and KMM4 are close both spatially and in velocity, as are KMM5 and KMM6, it is possible 
that they are part of the same structures. We therefore combined the inputs for the four substructures into KMM(3+4) and 
KMM(5+6) and re-ran the KMM algorithm on the 5 substructures. The results for KMM1, KMM2 and KMM7 were very similar to 
those found using 7 partitions, with similar galaxies being allocated to the combined KMM(3+4) and KMM(5+6) systems as
were allocated  when considering them as separate entities, and the overall correct allocation estimator for both 
the 5 and 7 substructure analyses being $98\%$ and $97.9\%$, respectively. It appears the algorithm does not favor a 7 
structure partition over a 5 structure partition, so we used the simpler decomposition and assumed the 5 substructure 
partition is the correct one. We tested the assumption of Gaussianity for the velocity distribution of each substructure
using a K-S test and found no evidence for significant deviations from Gaussianity. The results of the allocations to the 
different substructures are plotted in the top panel of Figure~\ref{kmmplot}, where we have color-coded the symbols to 
match the respective velocity distributions plotted in the bottom panel of Figure~\ref{kmmplot}.

\begin{deluxetable*}{cccccccc}
\tabletypesize{\scriptsize}
\tablecolumns{8}
\tablewidth{0pc}
\tablecaption{Results of the KMM analysis for partitioning the data into 7 ({\it top}) and 5 ({\it bottom})
substructures. See Figure~\ref{kmmplot} for the spatial and velocity distributions of the 5 partition results.\label{kmmtable}}
\tablehead{
\colhead{}&\multicolumn{3}{c}{Initial Inputs}& \multicolumn{4}{c}{KMM Outputs}\\
\cline{1-8}\\
\colhead{Group} &  \colhead{($\overline{x},\overline{y}, \overline{v}$)} &\colhead{($\sigma_{x}, \sigma_{y}, \sigma_{v}$)} 
&\colhead{$N_{gal}$}&\colhead{$(\overline{x},\overline{y}, \overline{v})$} &\colhead{($\sigma_{x}, \sigma_{y}, \sigma_{v}$)}
& \colhead{$N_{gal}$}&\colhead{Rate ($\%$)}\\
\cline{1-8}\\
\multicolumn{8}{c}{7 substructures}}
\startdata
KMM1&(138, 388, 444)&(103, 64, 175)&9&(150,  383,  432)&(145,  121,  167)&12&100\\
KMM2&(-1125, 550, -1816)&(320, 283, 318)&11&( -890,  346, -1887)&(419,  224,  254)&14&100\\
KMM3&(-431,-444, -996)&(62,134, 221)&7&( -462, -427, -1008)&(44,  139,  160)&6&100\\
KMM4&(-869,-63, -607)&(134, 116, 383)&12&( -924, -58, -672)&(146,  120,  257)&12&100\\
KMM5& (1563, 513,-642)&(200, 139, 233)&11&(1650,  533, -769)&(180,  132,  208)&10&98\\
KMM6& (2150,  1213, -236)&(155, 249, 170)&9&(2037,  1429, -317)&(401,  459, 220)&15&96\\
KMM7&(0, 0, 2)&(1236, 979, 738)&262&( -96,  93,  133)&(1211, 1016, 682)&252&98\\
\cutinhead{5 substructures}
KMM1&(138   388, 444)&(103, 64, 175)&9&(151,  382,  432)&(145,  121,  166)&12&100\\
KMM2&(-1125, 550, -1816)&(320, 283, 318)&11&(-890,  346, -1886)&(419,  224, 255)&14&100\\
KMM(3+4)&(-720, -200, -777)&(239,223, 321)&19&(-684, -249, -867)&(304,  264, 211))&16&94\\
KMM(5+6)&(1884, 849, -608)&(323, 421, 246)&20&(1831, 900, -753)&(348, 452, 328)&17&99\\
KMM7& (0, 0, 2)&(1236, 979, 738)&262&(-32, 138, 127)&(1250, 1028, 663)&262&98\\
\enddata
\tablecomments{The units of $\overline{x},\overline{y},\sigma_x$ and $\sigma_y$ are kpc, and the units of $\overline{v}$ 
and $\sigma_v$ are \kms.}
\end{deluxetable*}

\section{Merger Scenario}\label{mergescen}

Both the X-ray and optical analyses give clear indications that Abell~1201 hosts multiple substructures. In this section
we first give a qualitative scenario explaining the appearance of the X-ray structure and cold fronts through simple 
interpretations of the optical and X-ray observations combined with hydrodynamic simulations of 
\citet{poole2006} and \citet{ascasibar2006}. Second we use two-body analytic models to determine which of the 
substructures are bound to the main cluster.

\subsection{KMM1, the north-west X-ray excess and the formation of the cold fronts}\label{kmm1scen}

Figure~\ref{optxraycont} shows an SDSS $r$-band image of the central regions of Abell~1201 with X-ray contours overlaid, 
along with regions showing the cluster members and KMM1 allocations. Clearly, KMM1 is coincident with the excess X-ray 
emission which is probably the remnant of the gas core of KMM1. The morphology of the X-ray emission indicates the 
remnant KMM1 core is breaking up, and there appears to be a tail pointing towards the main cluster, although it is 
difficult to disentangle the cluster and subclump emission. The positioning of the cold fronts on opposite sides of the 
cluster center, along the direction to the north-west clump, suggests motion of the cluster core in this direction. This 
evidence points to a scenario where KMM1 has made its closest approach to the core of Abell~1201 and is traveling 
outwards toward the north-west. Projection effects make it difficult to know on exactly which plane the merger is occurring, 
although  the small radial velocity offset between KMM1 and Abell~1201 and the location of the cold fronts suggest the 
majority of the subcluster motion is in the plane of the sky. Since the core of Abell~1201 is not completely disrupted 
and is coincident with the dominant cluster galaxy, which presumably lies at the cluster potential minimum, it is unlikely
KMM1 passed directly through the cluster core \citep{poole2008}. The low velocity dispersion and compact galaxy 
distribution of KMM1 suggest we are seeing the surviving central region of a once larger cluster which has been stripped 
of its outer members due to the tidal effects of the main Abell~1201 cluster potential.

\begin{figure*}
\begin{center}
    {\includegraphics[angle=0,width=0.3\textwidth]{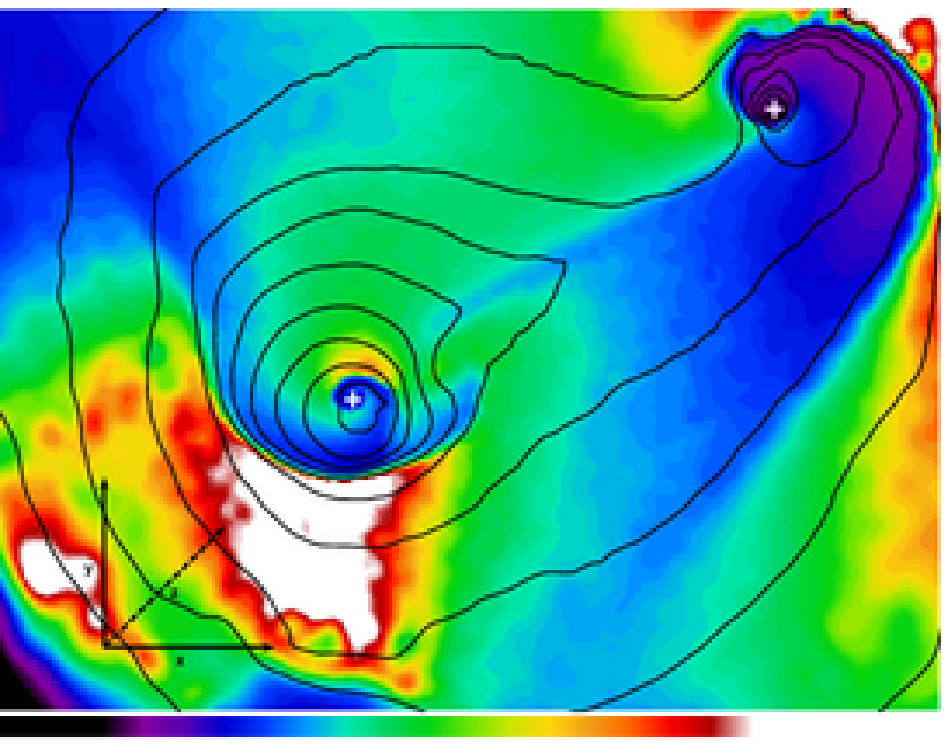}}
    {\includegraphics[angle=0,width=0.3\textwidth]{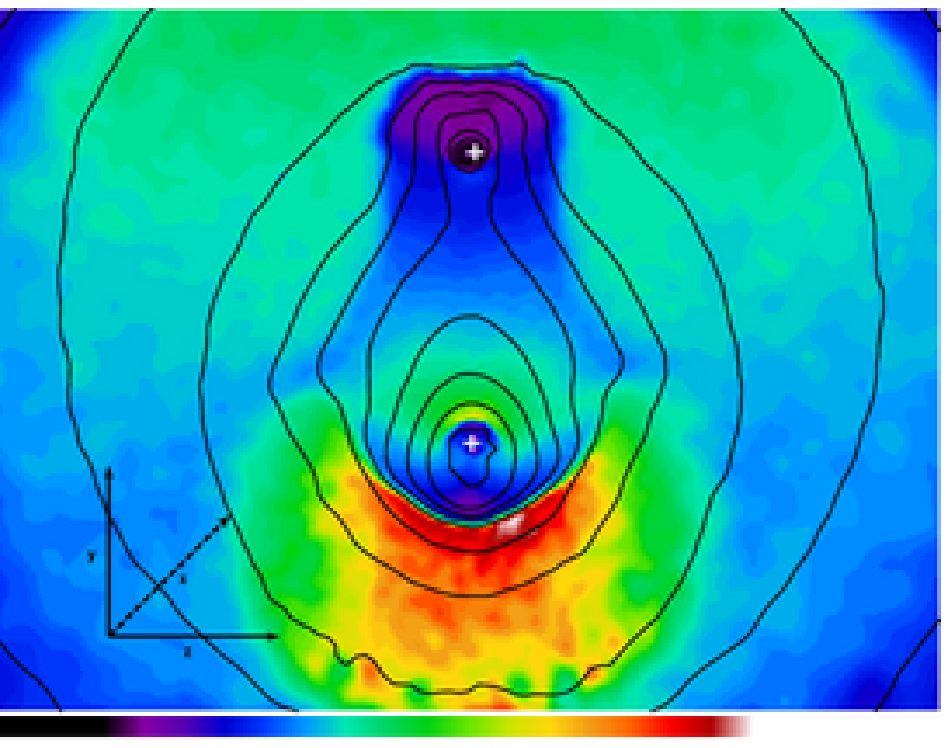}}
    {\includegraphics[angle=0,width=0.3\textwidth]{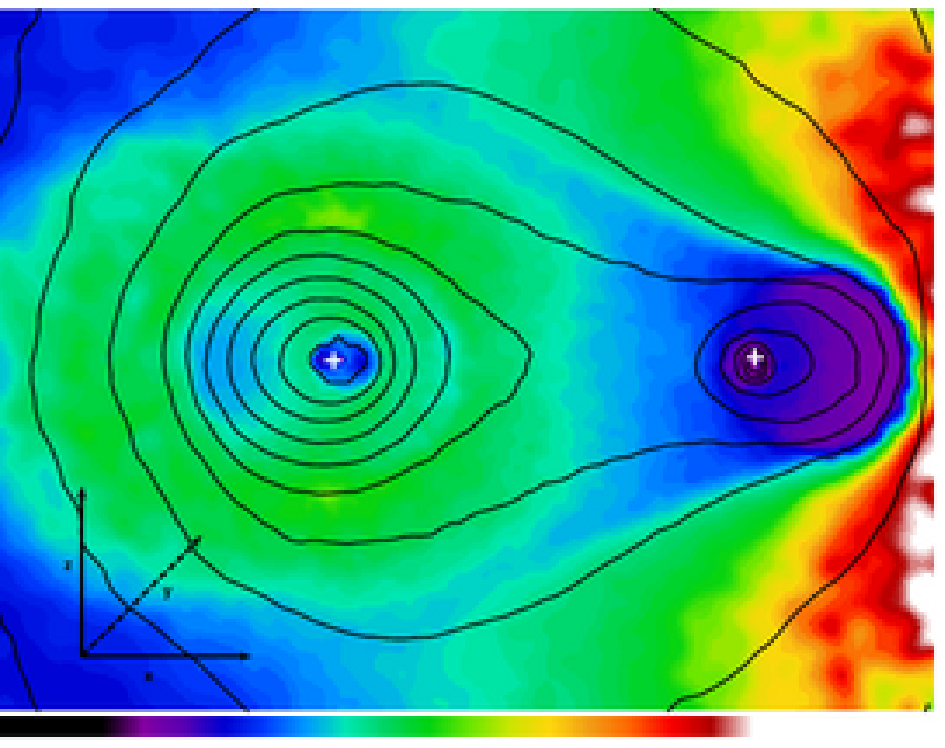}}
\end{center}
  \caption{Snapshot temperature maps of the 3 to 1 mass ratio, offset ($r_{min}=360$\,kpc) merger simulation from
\citet{poole2006} taken 0.7\,Gyrs after pericentric passage. The {\it left} panel shows the merger viewed from a vantage 
point where the line of sight is perpendicular to the merger orbital plane, while for the {\it middle} panel the line of 
sight is along the x-axis and the {\it right} panel the line of sight is along the y-axis. The vectors on the lower left
of each panel show this, and the length of each vector has physical size 500\,kpc. The color scale runs from black to white
with black showing the lowest temeperatures, and white the highest (on an arbitrary temperature scale). X-ray surface
brightness contours are overlaid and the white crosses show the positions of the peak in the dark matter density for the
primary and secondary clusters.}
\label{sim_imgs}
\end{figure*}

The south-east cold front appears to be connected to the core of the main cluster, as evidenced by the significant 
residuals extending from the core to the cold front, and also by the finger of cold ($\sim 4\,{\rm keV}$) gas joining the 
core and the cold front seen in the temperature map. This suggests the gas causing the cold front was once part of the 
core and has been displaced by the merger. It is possible that the cold fronts we see are the result of spirals of 
intertwined low and high entropy gas, as seen in both simulations \citep{ascasibar2006,poole2006} and observations 
\citep{fabian2006,clarke2004}, with the rotation axis of the spirals roughly in the plane of the sky. This scenario is
capable of producing several features observed in Abell~1201: two cold fronts at different distances on opposite sides of 
the core, the coincidence of the X-ray core with the dominant cluster galaxy and the region of hot gas on the north-west 
side of the cluster core which was probably heated to its observed temperature by compression or shock heating caused by 
the merger with KMM1. To illustrate this we present Figure~\ref{sim_imgs} which shows three temperature maps from the 
offset ($r_{min}=360$\,kpc), 3 to 1 mass ratio merger simulation of \citet{poole2006} generated using emission 
weighted temperatures integrated along a 3.5\,Mpc line-of-sight. There are 3 projections each at 0.7\,Gyrs 
after pericentric passage with one projection viewing along an axis perpendicular to the plane of the merger orbit 
($x - y$) and two projections orthogonal to this. This makes apparent the effect of projection -- the spiral type 
structure produces cold fronts observable over a wide range of viewing angles. Qualitatively, the map shown in the middle
panel of Figure~\ref{sim_imgs} reproduces the features of Abell~1201 noted above, although the secondary core appears 
farther from the cluster center than that in Abell~1201. If the orbit of the secondary core is close to our line-of-sight
and the core is close to turn around, this might be resolved by projecton effects.  However, the simulation illustrated 
has generic initial conditions, not specific to Abell~1201.  In particular, it is possible that a closer core passage 
could produce the structure observed in Abell~1201 without appealing to fortuitous viewing conditions.  The fact that an 
idealized simulation reproduces the majority of the observed features supports the proposed merger scenario.  Future
simulations specific to Abell~1201 will determine which of these possibilities can better explain its observed features.

It appears Abell~1201 is an excellent example of a ``sloshing'' type cold front cluster, where the perturber is still 
clearly visible in the X-ray and optical observations (in the form of KMM1), similar to Abell~1644 
\citep[see Figure 17][]{markevitch2007}. Thus, Abell~1201 is an excellent candidate for follow up detailed simulations 
in order to derive an accurate picture of exactly how the system has evolved to its current state, and whether it will 
evolve into a relaxed looking cluster harboring cold fronts with no discernable perturber \citep[eg.,  RXJ1720.1+2638, 
MS1455.0+2232 or Abell~2029][]{markevitch2007,mazzotta2008}

\subsection{Two-body merger dynamics}

In this section we apply the two-body dynamical analysis first implemented by \citet{beers1982} to the substructures 
detected in \S\ref{3dstructures} with the aim of determining which of the substructures are bound to the main cluster, and
to give an initial idea of the internal dynamics of the cluster. The model allows estimation of the probability that a 
substructure is unbound and lying close to the main cluster along the line of sight, and also allows the calculation of 
the probability that a substructure is bound and collapsing or bound and expanding. 

The accuracy of the results given by the model rely critically on the following assumptions: the orbits are radial, the 
cluster masses are concentrated into a point at the respective centers, the clusters had zero initial separation at t=0, 
and they are moving apart or coming together for the first time. These assumptions are extremely unlikely to hold true 
for KMM1, which probably resides well within the cluster virial radius, deep within the main cluster potential well (where 
dynamical friction, tidal forces and angular momentum all become significant) and is apparently currently heading to the 
north-west after pericentric passage. Thus, we do not present results from the two-body analysis for KMM1, and defer quantitative
analysis of the KMM1 merger until detailed simulations are available, noting only that given the observations 
presented here, it appears KMM1 and Abell~1201 form a bound system. 

As inputs, the model requires information about the projected spatial 
separation, $R_p$, the line of sight velocity difference, $V_r$, and the total mass of the system, returning possible 
solutions for $\alpha$ (the angle between the line joining the two clusters and the line of sight), the total mass 
required to bind the system, and the true 3-D spatial separation, $R$, and velocity difference, $V$. The parametric 
solutions to the equations of motion for 
bound radial orbits are
\begin{equation}\label{boundV}
V={{V_R} \over {\rm sin \alpha}} = {\left({2GM} \over {R_m} \right)^{1/2}} {{\rm sin \chi} \over {(1-\rm cos \chi)}},\\
\end{equation}
\begin{equation}\label{boundt}
t = {\left ( {R_m^3} \over {8 G M} \right)^{1/2}} \left (\chi - \rm sin \chi \right),
\end{equation}
\begin{equation}\label{boundR}
R= {{R_p} \over {\rm cos \alpha}} = {{R_m} \over {2}} \left ( 1- \rm cos \chi \right),
\end{equation}
where $R_m$ is the maximum separation of the two clusters at turn around, $M$ is the total mass of the system, and $\chi$ 
is the developmental angle which varies from $0<\chi<2\pi$ with $\chi=0,2\pi$ being the stages of the orbit corresponding 
to zero spatial separation. It is also possible to solve for the unbound case, where the parametric solutions are
\begin{equation}\label{unboundV}
V={{V_r} \over {\rm{sin} \alpha}} = {V_{\infty}} {{\rm{sinh} \chi} \over {(\rm{cosh} \chi -1)}},\\
\end{equation}
\begin{equation}\label{unboundt}
t = { {G M} \over {V_{\infty}^3}} \left (\rm{sinh} \chi - \chi \right),
\end{equation}
\begin{equation}\label{unboundR}
R= {{R_p} \over {\rm{cos} \alpha}} = {{G M} \over {V_{\infty}^2}} \left (\rm{cosh} \chi -1\right),
\end{equation}
where $V_{\infty}$ is the asymptotic expansion velocity. 

\begin{figure*}
{\includegraphics[angle=-90,width=1.\textwidth]{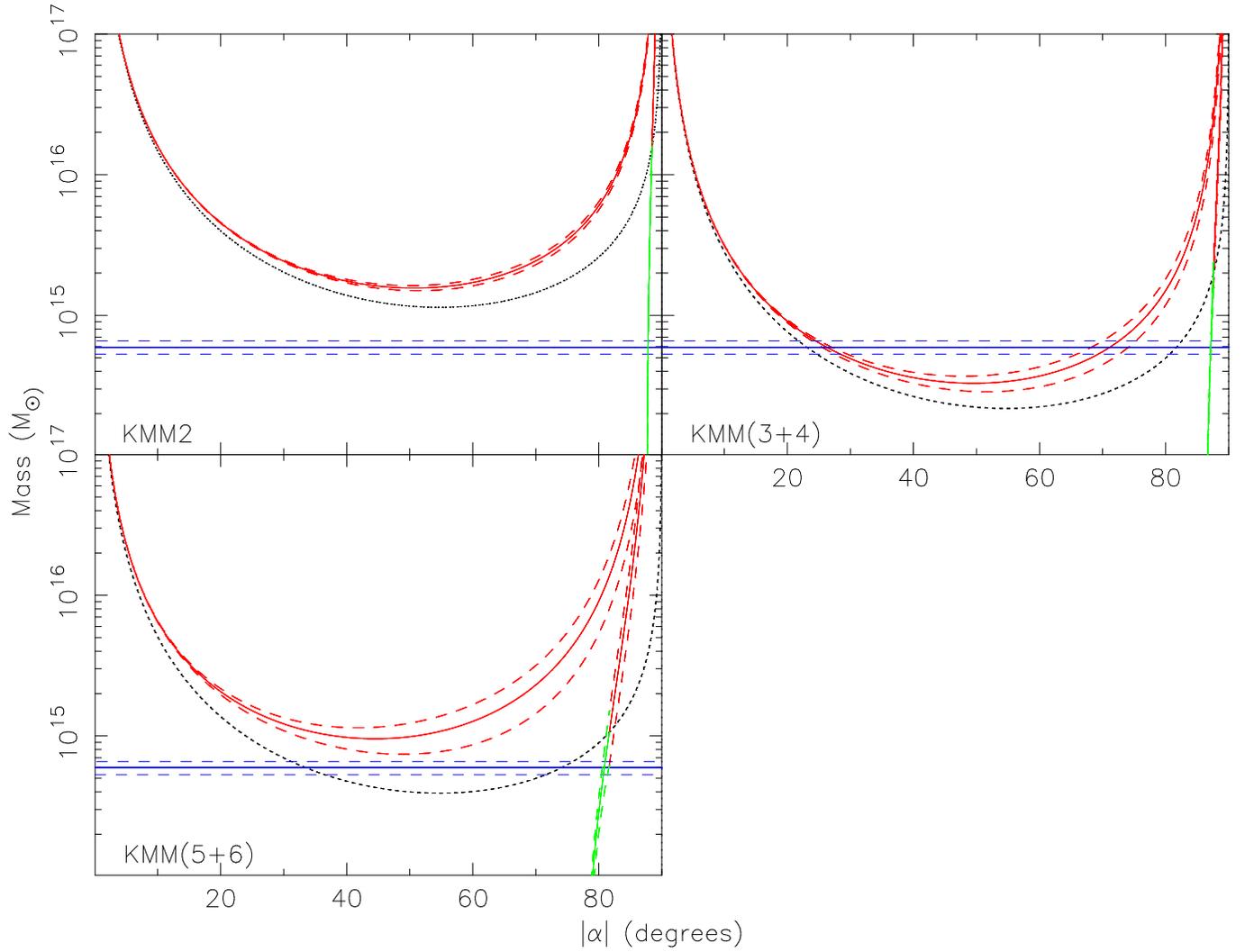}}
\caption{Binding mass as a function of $\alpha$ ({\it orange} and {\it green} curves). The {\it orange} curves correspond 
to bound solutions, while the {\it green} curves  represent unbound solutions. The {\it dashed} curves show the $1\sigma$ 
errors. The {\it blue} line shows the mass derived using galaxies allocated to KMM7 and the {\it dashed blue} lines 
indicate the upper and lower $1\sigma$ errors. The {\it small black dashed} line delineates the bound and unbound regions 
according to the Newtonian binding criterion.}
\label{twobodyplots}
\end{figure*}

For the bound case, combining equations \ref{boundV}, \ref{boundt} and \ref{boundR} gives
\begin{equation}
\rm{tan} \alpha = {{V_r t} \over {R_p}} {{(1-\rm{cos} \chi)^2} \over {\rm{sin} \chi (\chi - \rm{sin} \chi)}},
\end{equation}
and, similarly, we obtain
\begin{equation}
\rm{tan} \alpha = {{V_r t} \over {R_p}} {{(\rm{cosh} \chi -1)^2} \over {\rm{sinh} \chi ( \rm{sinh} \chi - \chi)}}.
\end{equation}
for the unbound case. Assuming  $t=11.38$\,Gyrs, i.e. the age of the Universe at $z=0.168$ in our assumed 
cosmology, for values of the parameter $\chi$ in the range 0 to $2\pi$ these equations determine possible solutions for 
$\alpha$. Given that the mass of the system is 
the least well constrained of the input parameters, we solved for $M$ as a function of $\alpha$ using as input 
$V_r$ and $R_p$ for each group in the KMM analysis. Here $R_p$ was calculated as the distance from the KMM centroid 
to the central dominant galaxy, rather than the KMM7 center, which is slightly offset from the central dominant galaxy 
position. $V_r$ is the radial velocity offset of the KMM partition of interest, taken with respect to the biweight 
location of peculiar velocities in KMM7, where either the median (for KMM groups with $N_{gal}<15$) or the biweight 
locator ($N_{gal}>15$) was used to determine the group's peculiar velocity. We plot the possible solutions of $M$ as a 
function of  $\alpha$ in Figure~\ref{twobodyplots}.

Determination of solutions for the orbits of the systems requires knowledge of the total mass of the system. Given
the velocity dispersions of the substructures in Abell~1201 are much smaller than that of the main cluster, it is 
reasonable to assume that when compared to the mass of the main Abell~1201 cluster the contribution of the masses of the 
subclumps is negligible. Hence we assumed the total system mass was equivalent to the virial mass of Abell~1201. We used 
the methodology of \citet{girardi1998} for the determination of the mass
\begin{equation}
M = M_{vir}- C = {{3 \pi} \over {2}} {{\sigma_v^2 R_{PV}} \over {G}} - C,
\end{equation}
where $C$ is the surface term correction accounting for the lack of coverage for the entire cluster 
(i.e. out to the turn around radius). Here we took  $C=0.19M_{vir}$, which is the median value derived in 
\citet{girardi1998}, and $\sigma_v=665\pm32$\kms\, which is the biweight scale estimator derived for galaxies assigned to 
KMM7 and
\begin{equation}
R_{PV} = {{N_{vir}(N_{vir}-1}) \over {\sum_{i=j+1}^{N_{vir}} \sum_{j=1}^{i-1} R_{ij}^{-1}}},
\end{equation}
which was determined within $r_{200}=0.17\sigma_v/H(z)=1.5$\,Mpc \citep{carlberg1997} where $N_{vir}=154$ is the 
number of KMM7 members within $r_{200}$, and $R_{ij}$ is the projected distance between two galaxies. We determined 
$R_{PV}=1.5\pm 0.1$\,Mpc and $M= 5.9(\pm0.6)\times10^{14} \rm{M}_{\odot}$ where the error in $R_{PV}$ was determined 
using the Jackknife technique, and standard error propagation was used to derive the error in the mass. 

The possible solutions for the orbits are given by the intersection of the line representing the Abell~1201 virial mass 
estimate with the curve for $M$ as a function of $\alpha$ in Figure~\ref{twobodyplots}. For each solution, we present 
the associated probability relative to the other solutions by considering the range of possible $\alpha$ given by the upper
and lower values, $\alpha_U$ and $\alpha_L$, respectively, derived from the error bounds on each curve, and assuming that 
each individual solution is equally probable. The relative probability is
\begin{equation}
p_{rel}={{\left(\int_{\alpha_L}^{\alpha_U} \rm{cos} \alpha\, \rm{d}\alpha \right)} \over {\sum p_{rel}}},
\end{equation}
for the solution of interest, and 
\begin{equation}
\sum p_{rel}=p_{BO}+p_{UO}+p_{BI_a}+p_{BI_b},
\end{equation}
where $p_{BO},\, p_{UO},\, p_{BI_a},\, p_{BI_b}$ are the $\int_{\alpha_L}^{\alpha_U} \rm{cos} \alpha\, \rm{d}\alpha$ values for
the bound outgoing, unbound outgoing, first bound incoming and second bound incoming solutions, respectively 
\citep{brough2006}. We present the results in Table \ref{twobodyresults}.

We also determined the probability the subclumps are bound using the Newtonian binding criterion, where a two-body system 
is bound if the kinetic energy is less than or equal to the potential energy. The criterion is given by \citet{beers1982}
as
\begin{equation}
V_r^2R_p \leq 2 GM\, \rm{sin}^2 \alpha\, \rm{cos} \alpha.
\end{equation}
This curve is also plotted in Figure~\ref{twobodyplots}, and the probability is  
$P_{bound}=\int_{\alpha_L}^{\alpha_U} \rm{cos} \alpha\, \rm{d}\alpha $, where  $\alpha_L$ and $\alpha_U$ are determined from
the intersections of $M_{vir}$ and the curve representing the Newtonian criterion. We present the results in Table 
\ref{twobodyresults}.

\begin{deluxetable*}{cccccccc}
\tabletypesize{\scriptsize}
\tablecolumns{8}
\tablewidth{0pc}
\tablecaption{Probabilities that the detected substructures are bound to the Abell~1201 cluster, derived using 
the two-body dynamical analysis and the Newtonian Criterion (see text).\label{twobodyresults}}

\tablehead{\colhead{Subclump} & \colhead{$V_r$} & \colhead{$R_{P}$}  & \colhead{$P_{BI_a}$} & \colhead{$P_{BI_b}$}
&\colhead{$ P_{BO}$} & \colhead{$P_{UO}$} & \colhead{$P_{bound}$}\\
&\colhead{(\kms)}&\colhead{(Mpc)}& \colhead{(per cent)}&\colhead{(per cent)}&\colhead{(per cent)}
&\colhead{(per cent)}&\colhead{(per cent)}\\}
\startdata
KMM2 & $-1989\pm46$& 0.95 &-&-&-&          100& -\\
KMM(3+4)&$-996\pm72$&0.73&62&38&-&$<<1$&60\\
KMM(5+6)& $-798\pm113$&2.04&-&-&$<<1$&100&41\\
\enddata
\end{deluxetable*}

Considering the Newtonian criterion and two-body solutions, we find that there is a high probability of KMM(3+4) being 
bound to Abell~1201, with two solutions which are, roughly speaking, as probable as each other. The most probable 
solution is KMM(3+4) is bound and incoming with $\alpha=26$\,degrees, $R=0.8$\,Mpc and $V=-2252$\kms, whilst the slightly 
less probable solution is also bound and incoming with $\alpha=71$\,degrees, $R=2.3$\,Mpc and $V=-1052$\kms. If KMM1 is 
outbound to the north-west, as proposed in \S~\ref{kmm1scen}, then it is possible that KMM(3+4) consists of tidal debris
stripped from KMM1 during is approach to pericentric passage from the south-east. Further evidence for a tidal origin 
comes from the spatial distribution of galaxies, where KMM(3+4) appears as two conglomerations in 
Figure~\ref{twobodyplots}, although simulations are required to confirm this.

In the case of KMM2, we find only unbound outgoing solutions with the most probable solution giving $\alpha=-88$\,degrees 
(i.e. it's motion is aligned at 2 deg to the line of sight), $R=24.8$\,Mpc and $V=1990$\kms. Thus it is likely 
that KMM2 is a group of galaxies lying in the foreground moving away from Abell~1201, and not physically associated. 
KMM(5+6) has a best fitting unbound outgoing solution, although within the errors this may be bound outgoing, with 
$\alpha=-81$\,degrees, $R=12.9$\,Mpc and $V=808$\kms. Given the Newtonian criterion allows KMM(5+6) to be bound 
and that it is possible we have underestimated the mass within the radius at which KMM(5+6) lies, it is entirely 
plausible that we have also underestimated the binding probability for KMM(5+6).

The two body analysis presented here has allowed the determination of which substructures are most likely bound to the
main cluster. This gives an initial understanding of which substructures are important in understanding the internal 
dynamics of Abell~1201, which will be further modeled in future simulations and presented in a forthcoming paper.

\section{Summary and Conclusions}\label{discussion}

We have presented an analysis of Abell~1201 using both \chan\, X-ray data and optical spectroscopic data
obtained using the AAT/AAOmega and MMT/Hectospec MOS. The X-ray analysis reveals the following:

\begin{enumerate}

\item An elliptical morphology with two surface brightness discontinuities positioned roughly on an axis joining the 
main cluster with a clump of excess emission to the north-west. The density, temperature and pressure jumps across these 
discontinuities are consistent with cold fronts. 

\item A residual map obtained by subtracting a double beta model from the surface brightness distribution shows 
significant residuals at the position of the north-west clump. It shows significant residuals extending from the
cluster core to the south-east cold front. 

\item A temperature map shows that south-east residual is coincident with a finger of cold
gas which also extends from the cluster core and terminates on the inner side of the front. 

\item There is also a significantly hotter region on the north-west side of the inner cold front between the main cluster core 
and the subclump further to the north-west. 

\end{enumerate}

The optical MOS analysis reveal:

\begin{enumerate}

\item From 321 cluster member spectra the cluster redshift is z=0.1673 and the velocity dispersion is 
$778$\kms. 

\item The velocity distribution is not significantly non-Gaussian, despite the clearly disturbed 
nature of Abell~1201 evidenced by the X-ray data.

\item Combining the peculiar velocity and spatial information reveals significant localized velocity 
substructure.

\item Using the KMM method of \citet{ashman1994} the substructure can be partitioned into 5 distinct clumps - the main 
Abell~1201 cluster with 262 members (KMM7), an infalling subgroup $\sim 410$\,kpc to the 
north-west coincident with the X-ray excess containing  12 members (KMM1), a subgroup $\sim 950$\,kpc to the NE with 14 
members (KMM2), a subgroup $\sim 730$\,kpc to the south-east with 16 members (KMM3+4) and a group $\sim 2$\,Mpc to the north-west with 
15 members (KMM5+6). 

\item Application of the two-body dynamical analysis of \citet{beers1982} reveals KMM2 is likely an unbound foreground 
group, whilst it is also possible KMM(5+6) is unbound and KMM(3+4) has a high probability of being bound to the main 
cluster.
\end{enumerate}

The analysis presented here supports the view that the cold fronts in Abell~1201 are a direct result of merger activity. 
Our X-ray and optical analysis combined with the simulation of \citet{poole2006} indicate that the KMM1 subcluster has 
come from the south-east and is currently heading towards the north-west after an offset passage past the main cluster
core. Disturbance caused by the passage of this subcluster has set off sloshing of the cool gas in 
the core of the main cluster, which produced two concentric cold fronts around the main core. The value of the use of 
multiple methods of merger detection for analyzing a clusters dynamical state is also confirmed in this study.

The next step in our study of cold front clusters will be to utilize radio data in combination with the optical 
spectroscopy presented here to study the effects of the merger on the galaxy population, and search for signs of 
non-thermal radio halo/relic emission. Deeper X-ray observations will help place tighter constraints on the parameters 
derived here, whilst allowing a more detailed analysis of the complex X-ray structure. 

\section{Acknowledgments}

We thank David Woods for useful discussions and sanity checks. We are grateful to Will Saunders and the staff at the 
Anglo-Australian Observatory for their support during AAT observations. Observations reported here were obtained at the 
MMT Observatory, a joint facility of the Smithsonian Institution and the University of Arizona. We thank the MMT 
operators and queue-schedule mode scientists for their help during observations and the staff at the 
Harvard-Smithsonian Center for Astrophysics Telescope Data Center for reducing the Hectospec data.

This research has made use of software provided by the Chandra X-ray Center (CXC) in the application packages CIAO, 
ChIPS, and Sherpa and also of data obtained from the Chandra archive at the NASA Chandra X-ray center 
(http://cxc.harvard.edu/cda/). This research has made use of the NASA/IPAC Extragalactic Database (NED) which is operated 
by the Jet Propulsion Laboratory, California Institute of Technology, under contract with the National Aeronautics and 
Space Administration.

MSO was supported by an Australian Postgraduate Award, and acknowledges the hospitality of the Harvard-Smithsonian Center
for Astrophysics where a portion of this study was undertaken. We acknowledge the financial support of the Australian 
Research Council (via its Discovery Project Scheme) throughout the course of this work. PEJN was supported by NASA 
grant NAS8-01130.

\label{lastpage}
\end{document}